\documentclass[11pt,draftcls,onecolumn]{IEEEtran}
\usepackage[english]{babel}
\usepackage[utf8]{inputenc}
\usepackage{amsmath}
\usepackage{amssymb}
\usepackage{bbm}
\usepackage{algorithm,algcompatible,amsmath}
\usepackage{titling}
\newtheorem{theorem}{Theorem}[section]

\newtheorem{lemma}[theorem]{Lemma}

\usepackage{graphicx}
\usepackage{amstext}
\usepackage{amsmath}
\usepackage{amsfonts,amssymb}
\usepackage{amsmath,tabu}
\usepackage{amssymb}
\usepackage[dvips]{epsfig}
\usepackage{epsfig}
\usepackage[dvips]{graphicx}
\usepackage{multirow}
\usepackage{multirow} 
\usepackage[utf8]{inputenc}
\usepackage[english]{babel}
\usepackage{caption}
\usepackage{subcaption}
\usepackage{float}
\usepackage[nospace,noadjust]{cite}

\def\bth{{\boldsymbol{\theta}}}

\def\bGam{{\boldsymbol{\Gamma}}}

\def\bPsi{{\boldsymbol{\Psi}}}
\def\bLam{{\boldsymbol{\Lambda}}}
\def\blam{{\boldsymbol{\lambda}}}
\def\b1{{\boldsymbol{1}}}
\def\c1{{\textcircled{a}}}

\def\bn{{\boldsymbol{n}}}

\def\br{{\mathbf{r}}}
\def\bs{{\boldsymbol{s}}}

\def\bu{{\boldsymbol{u}}}

\def\bw{{\boldsymbol{w}}}
\def\bx{{\boldsymbol{x}}}
\def\by{{\mathbf{y}}}

\def\bA{{\mathbf{A}}}
\def\bB{{\boldsymbol{B}}}
\def\bC{{\boldsymbol{C}}}
\def\bD{{\boldsymbol{D}}}
\def\bE{{\boldsymbol{E}}}
\def\bF{{\mathbf{F}}}
\def\bG{{\boldsymbol{G}}}

\def\bI{{\mathbf{I}}}

\def\bP{{\mathbf{P}}}
\def\bQ{{\boldsymbol{Q}}}
\def\bR{{\mathbf{R}}}

\def\bT{{\mathbf{T}}}
\def\bU{{\boldsymbol{U}}}
\def\bV{{\boldsymbol{V}}}
\def\bW{{\boldsymbol{W}}}
\def\bX{{\boldsymbol{X}}}
\def\bY{{\boldsymbol{Y}}}
\def\bZ{{\boldsymbol{Z}}}

\def\bzero{{\boldsymbol{0}}}

\def\hatbX{{\boldsymbol{\hat{X}}}}
\providecommand{\keywords}[1]{\textbf{\textit{Index terms---}} #1}
\makeatletter
\def\thanks#1{\protected@xdef\@thanks{\@thanks
        \protect\footnotetext{#1}}}
\makeatother
\begin{document}
\title{New Methods for MLE of Toeplitz Structured Covariance Matrices with Applications to RADAR Problems}
\author{Augusto Aubry, Prabhu Babu, Antonio De Maio,  and Rikhabchand Jyothi 
\thanks{Augusto Aubry and Antonio De Maio  are with the Department of Electrical Engineering and Information
Technology, Universita degli Studi di Napoli ``Federico II”, DIETI, Via Claudio 21, I-80125 Napoli, Italy (E-mail:
augusto.aubry@unina.it, ademaio@unina.it). Prabhu Babu and Rikhabchand Jyothi are with CARE, IIT Delhi, New Delhi, 110016, India (E-mail: prabhubabu@care.iitd.ac.in, jyothi.r@care.iitd.ac.in)
}}
\date{}
\maketitle
\begin{abstract}
This work considers Maximum Likelihood Estimation (MLE) of a Toeplitz structured covariance matrix. In this regard, an equivalent reformulation of the MLE problem is introduced and two iterative algorithms are proposed for the optimization of the equivalent problem. Both the strategies are based on the Majorization Minimization (MM) framework and hence enjoy nice properties such as monotonicity and ensured convergence to a stationary point of the equivalent MLE problem. The proposed algorithms are also extended to deal with MLE of other related covariance structures, namely, the banded Toeplitz, Toeplitz-block-Toeplitz, low rank Toeplitz structure plus a scalar matrix (accounting for white noise), and finally Toeplitz matrices satisfying a condition number constraint. Through numerical simulations, it is shown that new methods provide satisfactory performance levels in terms of both mean square estimation error and signal-to-interference-plus-noise ratio.
\end{abstract}
\section{Introduction}
Estimation of the interference covariance matrix has diverse applications in radar signal processing such as direction
of arrival estimation, target detection, adaptive beamforming, and sidelobe canceller design (\cite{amf, stoica,app,farina}). In these situations, the interference covariance matrix is estimated 
from the secondary/training data, which are assumed target-free and collected from spatial and/or temporal returns corresponding to range cells close to the one of interest. The Sample Covariance Matrix (SCM), which is the unstructured Maximum Likelihood Estimate (MLE)  when the data follows a complex circular Gaussian distribution, does not always represent a good choice for the covariance estimate especially in the presence of a small number of training data and/or when mismatches in training data spectral properties occur \cite{scm1,scm2}. A well-known strategy often discussed in the open literature to improve the performance of a covariance estimator relies on the incorporation of some \emph{a priori} knowledge about its structure. For instance, in some radar applications, it is customary to suppose that data come from a  stationary Gaussian random process, leading to a  Toeplitz Structured Covariance (TSC) matrix. Leveraging this information, one can obtain (under the design conditions) a more reliable estimator as compared with the SCM \cite{toep}. Other than radar applications, the estimation of a TSC matrix is encountered in speech recognition \cite{speech}, spectral estimation \cite{stoica}, and in hyperspectral imaging \cite{imaging}.\\
So far several algorithms have been proposed for estimating a TSC matrix. Let us first discuss those for MLE. According to the Caratheodory parametrization (\cite{stoica,caratheodoryp1,caratheodoryp2}) a Toeplitz covariance matrix $\bT \in \mathbb{H}^{m \times m}$ can always be decomposed as
\begin{equation}\label{cp}
    \begin{array}{ll}
    \bT = {\bA}\tilde{\bP}{\bA}^{H}; \:\: [\tilde{\bP}]_{k,k}\geq 0
    \end{array}
\end{equation}
where  
\begin{equation}
\bA= 
\begin{bmatrix}
1 & 1 & \cdots & 1\\
e^{j\omega_{1}} & e^{j\omega_{2}} & \cdots & e^{j\omega_{r}} \\
\vdots  & \vdots  & \ddots & \vdots  \\
e^{j(m-1)\omega_{1}} & e^{j(m-1)\omega_{2}} & \cdots & e^{j(m-1)\omega_{r}}
\end{bmatrix}
\end{equation}
\begin{equation}
 \tilde{\bP} =
  \begin{bmatrix}
    \tilde{p}_{1} & 0&\dots&0 \\
    0 &  \tilde{p}_{2} & \dots & 0 \\
     \vdots & \vdots & \ddots & \vdots \\
    0&0&\dots& \tilde{p}_{r}
  \end{bmatrix}
\end{equation}
where $\omega_{i}$ and $\tilde{p}_{i}$, $i=1,2, \cdots,r \leq m$, denote some angular frequencies and their corresponding powers and $r$ indicates the rank of $\bT$. Capitalizing on this parametrization, Circulant Embedding (CE) of Toeplitz matrix (\cite{ce1,ce2,ce3}) can be used  to compute approximately the MLE of $\bT$. According to CE, a positive semidefinite $m \times m$ Toeplitz matrix is modeled as:
\begin{equation}\label{ce}
    \begin{array}{ll}
    \bT = \tilde{\bF}\bP\tilde{\bF}^{H}; \:\: \bP= {\textrm{diag}}([p_{1},p_{2},\cdots,p_{L}]), p_{k} \geq 0
    \end{array}
\end{equation}
where $\tilde{\bF} = [\bI_{m \times m} \: \bzero_{m \times L-m}]\bF$, $\bI_{m \times m}$ is the identity matrix of size $m \times m$, $\bzero_{m \times L-m}$ is the zero matrix of size $m \times L-m$, $\bF$ is the normalized DFT matrix of size $L \geq 2m-1$ and $\bP$ is a diagonal matrix of size $L \times L$ with diagonal elements $p_{k} \geq 0$. Therefore, the matrix $\bT$ is completely parametrized by the diagonal matrix $\bP$. Although estimating the Toeplitz covariance matrix using CE seems attractive, the representation in (\ref{ce}) is valid only for a subset of Toeplitz covariance matrices. This can be intuitively justified because the Caratheodory parametrization in (\ref{cp}) does not give restrictions on the frequencies spacing, while the CE in (\ref{ce}) strictly requires the frequencies to lie on the Fourier grid. Hence, for some Toeplitz matrices, the parametrization in (\ref{ce}) is only approximate. Based on CE, \cite{em1} and \cite{em2} proposed an iterative algorithm based on Expectation-Maximization (EM) for MLE of $\bT$. By modifying the M step in the EM procedure, in \cite{band} the technique has been extended to deal with the banded Toeplitz covariance case. In \cite{melt}, still leveraging CE framework, a Majorization Minimization (MM) based optimization, with faster convergence than the EM in \cite{em1} and \cite{em2}, has been introduced. In \cite{ip} a closed-form estimator is designed by invoking the extended invariance principle to deal with the Toeplitz constraint. Finally, in \cite{eastr}, an efficient approximation of a Toeplitz covariance matrix under a rank constraint has been proposed forcing the eigenvectors to be the same as those of the SCM whereas the Toeplitz constraint has been explicitly imposed while estimating the eigenvalues. Other than the MLE, several other alternative paradigms have been considered for the problem at hand. Recently, in \cite{Augusto} the Toeplitz structure is forced together with a condition number constraint via SCM projection onto a suitable constraint set.  Other geometric based approaches for the TSC estimate have also been proposed in \cite{geo1,geo2}.\\
In this manuscript, two iterative algorithms denoted as \textbf{A}lternating Projection Based \textbf{TO}eplitz Covariance \textbf{M}atrix Estimation 1 (ATOM1)  and ATOM2 are devised leveraging a suitable reformulation of the MLE problem and the MM framework. Both ATOM1 and ATOM2 involve the construction of a surrogate function together with its optimization. Specifically, the two procedures construct distinct surrogate functions and therefore solve different surrogate minimization problems. While ATOM1 addresses the surrogate minimization problem using the Alternating Direction Method of Multipliers (ADMM), ATOM2 handles it via alternating projection or Dykstra's algorithm. However, both the procedures directly estimate the Toeplitz covariance matrix without reparametrization via the CE. ATOM2 is also extended to include other constraints, such as an upper bound to the condition number and a low rank  plus white noise covariance structure. Finally, ATOM2 is modified to deal with banded Toeplitz and Toeplitz-block-Toeplitz structures. The major contributions of this paper can be summarized as follows: 
\begin{enumerate}
    \item{Two iterative algorithms ATOM1 and ATOM2 are proposed based on the MM framework to address MLE of a Toeplitz covariance matrix. Their computational complexities are thoroughly discussed. Also, the convergence of the procedures to a stationary point of the MLE problem is established.}
    \item{The extensions of ATOM2 to handle additional covariance structures, such as banded Toeplitz, Toeplitz-block-Toeplitz, and low rank Toeplitz term plus a scalar matrix are discussed, together with the possibility to account for a condition number constraint.}
    \item{The Cramer-Rao lower bounds for the MLE estimation of Toeplitz, banded Toeplitz, and Toeplitz-block-Toeplitz covariance matrices are derived}.
    \item{The proposed algorithms and their extensions are compared with some state-of-the-art procedures via numerical simulations using the Mean Square Error (MSE) and Signal-to-Interference-plus-Noise Ratio (SINR) (for case studies related to radar application) as performance metrics.}
\end{enumerate}
The organization of the paper is as follows. The MLE problem of Toeplitz covariance matrix for complex circular Gaussian observations is formulated in Section \ref{sec:2}. In Section \ref{sec:3}, ATOM1 and ATOM2 algorithms are proposed, along with a discussion on their computational complexity and implementation aspects. Also, their convergence properties are studied. At the end of the section, the extension of ATOM2 to handle additional constraints along with the Toeplitz requirement is discussed too. In Section \ref{sec:crlb}, the Cramer-Rao lower bounds for the MLE of Toeplitz, banded Toeplitz, and Toeplitz-block-Toeplitz covariance matrices are computed. In Section \ref{sec:4}, the proposed algorithms are compared with some state-of-the-art algorithms, and finally, concluding remarks are given in Section \ref{sec:5}.\\ 
\textbf{A}. Notation\\
Throughout the paper, {{bold}} capital and {{bold}} small letter denote matrix and vector, respectively. A scalar is represented by a small letter. The value taken by an optimization vector ${\bx}$ at the ${t^{th}}$ iteration is denoted by ${\bx_{t}}$. The notations $\mathbb{R}$, $\mathbb{R}^{m \times 1}$, $\mathbb{C}^{m \times 1}$, $\mathbb{R}^{m \times m}$, $\mathbb{C}^{m \times m}$, and
$\mathbb{H}^{m \times m}$ are used to represent the sets of real numbers, $m$ dimensional vectors of real
numbers, $m$ dimensional vectors of complex numbers, $m \times m$ matrices of real numbers, $m \times m$ matrices of complex numbers, and $m \times m$ Hermitian matrices, respectively. Superscripts $(\cdot)^{T}$, $(\cdot)^{*}$,  $(\cdot)^{H}$, and $(\cdot)^{-1}$ indicate the transpose, complex conjugate, complex conjugate transpose, and inverse, respectively. The trace and the determinant of a matrix $\bX$ are denoted by ${\rm{Tr}}(\bX)$ and $|\bX|$, respectively. The notation $[{\bX}]_{i}$ is used to represent the $i^{th}$ column of the matrix $\bX$. The $\otimes$ symbol indicates the Kronecker product while the gradient of a function $f$ is denoted by $\nabla f$. The  $\succeq$ symbol (and its strict form $\succ$) is used to denote the generalized matrix inequality: for any $\bX \in \mathbb{H}^{m \times m}$,  $\bX \succeq  0$ means that $\bX$ is a positive semi-definite matrix ($\bX \succ 0$ for positive definiteness). Besides, for any $\bX \in \mathbb{H}^{m \times m}$, ${\rm{eig}}(\bX)$ is the vector collecting the eigenvalues of $\bX$ with the maximum and the minimum eigenvalue indicated as $\lambda_{max}(\bX)$  and $\lambda_{min}(\bX)$, respectively. The Euclidean norm of the vector $\bx$ is denoted by ${\|\bx\|}_{2}$, $|\bx|$ indicates the element wise modulus of the vector $\bx$. The notation $\textbf{E}[\cdot]$ stand for statistical expectation. Finally, for any $\bX,\bY \in \mathbb{R}^{m\times m}$, $\rm{max}(\bX,\bY)$ refers to the matrix containing the element wise maximum between  $\bX$ and $\bY$.

\section{Problem Formulation}\label{sec:2}
Let us assume the availability of $n$ independent and identically distributed vectors $\{\by_{1}, \by_{2}, \cdots,\by_{n}\}$, where each $\by_{i}$ is of size $m$ and follows a $m$-variate complex circular Gaussian distribution with zero mean and covariance matrix $\bR$. The maximum likelihood covariance estimation problem can be formulated as  
\begin{equation} \label{eq:11}
\begin{array}{ll}
\underset{\bR \succ 0}{\rm minimize} \: f(\bR) {=}\dfrac{1}{n}\displaystyle\sum_{i=1}^{n}\by_{i}^{H}\bR^{-1}\by_{i} + \log|\bR| 
\end{array}
\end{equation}
If $n \geq m$, problem (\ref{eq:11}) has a unique minimizer with probability one which is given by the SCM, i.e., $\bR_{\textrm{SCM}} = \dfrac{1}{n}\displaystyle\sum_{i=1}^{n}\by_{i}\by_{i}^{H}$.  However, if the random process, from which each observation is drawn, is  stationary (at least in wide sense) then the covariance matrix also exhibits a Toeplitz structure which can be capitalized in the estimation process. By doing so, problem (\ref{eq:11}) becomes
\begin{equation} \label{eq:13}
\begin{array}{ll}
\textrm{MLE:}\quad\underset{\bR \in Toep, \bR \succ 0}{\rm minimize} \: f(\bR)
\end{array}
\end{equation}
where $Toep$ is used to denote the set of Toeplitz matrices of size $m \times m$. The above problem has two constraints: a structural constraint and a positive definite constraint. Even though the structural constraint is convex, the non-convexity of the objective function makes Problem (\ref{eq:13}) challenging to solve and no analytical solution seems to be available. In the following two iterative solution procedures for (\ref{eq:13}) are designed exploiting the MM principle. Briefly, the MM technique mainly consists of two steps - 1) Constructing a surrogate function $g(\bR|\bR_{t})$ (where $\bR_{t}$ is the estimate of $\bR$ at the $t^{th}$ iteration) for the objective function in (\ref{eq:13}) and 2) Minimizing the resulting surrogate problem at each iteration. For more details, \cite{MM1, MM2, conv, am} provide an in-depth discussion on MM based algorithms.
\section{Algorithms for Toeplitz covariance matrix estimation}\label{sec:3}
In this section, ATOM1 and ATOM2 are proposed to tackle the MLE problem of TSC matrix. Both leverage the MM principle (applied to an equivalent reformulation of the MLE problem) and differ in the way they construct and handle the surrogate minimization problem. ATOM1 solves the surrogate optimization using ADMM while ATOM2 tackles it using alternating projection or Dykstra's  algorithm. Hence the computational complexity and proof of convergence of the procedures are established. Finally, the extension of ATOM2 to deal with additional covariance constraints along with the  Toeplitz structure is provided. \\
To begin with, let us present the following two lemmas which pave the way for the development of ATOM1 and ATOM2.  
\begin{lemma} \label{lemma 1}
Consider the following two optimization problems
\begin{equation}\label{n0}\tag{$P_{1}$}
\begin{array}{ll}
\bR^{*}=\underset{\bR  \in Toep, \bR \succ 0}{\rm arg\:min} \: \dfrac{1}{n}\displaystyle\sum_{i=1}^{n}\by_{i}^{H}\bR^{-1}\by_{i}  + \log|\bR| 
\end{array}
\end{equation} 
\begin{equation}\label{n1}\tag{$P_{2}$}
\begin{array}{ll}
\bX^{*}=\underset{{\bX}  \in Toep, \bX \succ 0}{\rm arg\:min} \: \log|\bX|  \:\:\:
{\rm s.t.}\quad \dfrac{1}{n}\displaystyle\sum_{i=1}^{n}\by_{i}^{H}{\bX}^{-1}\by_{i} \leq 1
\end{array}
\end{equation}
Problems $(P_{1})$ and $(P_{2})$ are equivalent and $\bR^{*}$ can be obtained via $\bX^{*}$ using the following relationship
\begin{equation}\label{relation}
\bR^{*} = \dfrac{\bX^{*}}{m}
\end{equation}
\begin{IEEEproof}
The proof can be found in Appendix A. 
\end{IEEEproof}
\end{lemma}
Before proceeding with the next important lemma, it is worth pointing out that Lemma \ref{lemma 1} holds true even if the Toeplitz structural constraint in $(P_{1})$ and $(P_{2})$ is replaced by an arbitrary closed-cone set.
\begin{lemma}\label{lemma 0}
Given a concave differentiable\footnote{For a non-differentiable function, the inequality in (\ref{l0}) can be relaxed to  $f(\bX) \leq f(\bX_{t}) + {\rm{Tr}}\left(\bG(\bX_{t})^{H} (\bX-\bX_{t})\right)$, where $\bG(\bX_{t})$ is the subgradient of the concave function $f(\bX)$ at $\bX_{t}$. } function $f(\bX): \mathbb{H}^{m \times m} \rightarrow \mathbb{R}$, it can be majorized as
\begin{equation}\label{l0}
    \begin{array}{ll}
    f(\bX) \leq f(\bX_{t}) + {\rm{Tr}}\left(\nabla f(\bX_{t})^{H} (\bX-\bX_{t})\right)
    \end{array}
    \end{equation}
where $\bX ={\bX}_{t}$ is the value of $\bX$ at the $t^{th}$ iteration. The upper bound to $f(\bX)$ is linear and differentiable with respect to (w.r.t.) $\bX$.
\begin{IEEEproof}
Since $f(\bX)$ is a concave function w.r.t. $\bX$, (\ref{l0}) stems from linearizing $f(\bX)$ via its first order Taylor expansion.
\end{IEEEproof}
\end{lemma}
Leveraging Lemma \ref{lemma 1}, a minimizer of Problem ($P_{1}$), i.e., the MLE problem in (\ref{eq:13}), can be obtained by solving Problem ($P_{2}$) and using the relationship in (\ref{relation}). Therefore, in the following, let us focus on solving Problem ($P_{2}$). Now, since $\log|\bX|$ in (\ref{n1}) is a concave function w.r.t. $\bX$ \cite{boyd}, it can be majorized using Lemma \ref{lemma 0} to get the following surrogate function 
 \begin{equation}\label{ss}
\begin{array}{ll}
g(\bX|\bX_{t}) = \textrm{Tr}\left(({\bX}_{t})^{-1}\bX\right) \:{\rm s.t.}\quad \dfrac{1}{n}\displaystyle\sum_{i=1}^{n}\by_{i}^{H}{\bX}^{-1}\by_{i} \leq 1
\end{array}
\end{equation}
Therefore, at any iteration, given $\bX_{t}$, the MM method demands for the solution of  the following surrogate minimization problem
\begin{equation} \label{s1}
\begin{array}{ll}
\bX_{t+1} = \underset{\bX  \in Toep, \bX \succ 0}{\rm arg\: min} \: \textrm{Tr}\left(({\bX}_{t})^{-1}\bX\right) \:{\rm s.t.}\quad \dfrac{1}{n}\displaystyle\sum_{i=1}^{n}\by_{i}^{H}{\bX}^{-1}\by_{i} \leq 1
\end{array}
\end{equation}

This a convex optimization problem and can be cast as a SemiDefinite Program (SDP) (the interested reader may refer to Appendix B for details)
\begin{equation} \label{n5}
\begin{array}{ll}
 \underset{{\bX\in Toep}}{\rm minimize} \: \textrm{Tr}\left(({\bX}_{t})^{-1}{\bX}\right)\\
{\rm subject\ to}\:\: \boldsymbol{\hat{X}} = \bI \otimes \bX \succeq \dfrac{1}{n} \bar{\by}\bar{\by}^{H}
\end{array}
\end{equation}
where $\boldsymbol{\hat{X}}$ is a block diagonal matrix with the diagonal blocks equal to $\bX$ and $\bar{\by}$ is a vector obtained by stacking the vectors $\by_{1},\by_{2},\cdots,\by_{n}$. However, the computational complexity necessary to handle SDP using interior point methods is $\mathcal{O}(m^{3}n^{3})$ \cite{sdpcomplexity}.
In the following, an iterative algorithm based on ADMM is proposed to solve the surrogate minimization problem in (\ref{n5}). 

\subsection{ATOM1}\label{sec:3a}
The surrogate minimization problem in (\ref{n5}) is solved using ADMM \cite{admmmatrix1,admmmatrix2}. To do so, an auxiliary variable $\bU$ is introduced in (\ref{n5}) and the problem is re-written in the equivalent form
\begin{equation}\label{admm1}
\begin{array}{ll}
\underset{{\bX}  \in Toep,\: \bU}{\rm minimize} \:& \textrm{Tr}\left(({\bX}_{t})^{-1}{\bX}\right) \\
{\rm subject\ to} & \quad {\bI \otimes \bX} - \bU = \dfrac{1}{n} \bar{\by}\bar{\by}^{H} \\
&\quad \quad  \bU \succeq \boldsymbol{0} 
\end{array}
\end{equation}
The augmented Lagrangian for (\ref{admm1}) is
\begin{equation}
    \begin{array}{ll}
    L_{\rho}(\bX,\bU,\hat\blam) = \textrm{Tr}\left(({\bX}_{t})^{-1}{\bX}\right) + {\rm{Tr}}\left({(\hat{\blam})}^{T} \left((\bI \otimes \bX) - \bU -\dfrac{1}{n}\bar{\by}\bar{\by}^{H}\right)\right) + \dfrac{\rho}{2}\|(\bI \otimes \bX) -\bU -\dfrac{1}{n} \bar{\by}\bar{\by}^{H} \|_{F}^{2}
    \end{array}
\end{equation}
where $\rho>0$ is the penalty parameter and $\hat{\blam} \succeq 0$ is the Lagrange multiplier of size $mn \times mn$. The iterative steps of ADMM algorithm are
\begin{equation}\label{updateX}
\begin{array}{ll}
\bU^{t}_{k+1}=\underset{{\bU \succeq 0}}{\rm arg\: min}\: {\rm{Tr}}\left({(\hat{\blam}^{t}_{k})}^{T} \left((\bI \otimes \bX_{k}^{t}) - \bU\right)\right) +  \dfrac{\rho}{2}\|(\bI \otimes \bX_{k}^{t}) -\bU -\dfrac{1}{n} \bar{\by}\bar{\by}^{H} \|_{F}^{2}\\
\end{array}
\end{equation}
\begin{equation}\label{updateR}
\begin{array}{ll}
\bX^{t}_{k+1}=\underset{{\bX  \in Toep}}{\rm arg\:min}\: \textrm{Tr}\left(({\bX}_{t})^{-1}{\bX}\right) + {\rm{Tr}}\left({(\hat{\blam}^{t}_{k})}^{T}\left((\bI \otimes \bX)\right)\right) +  \dfrac{\rho}{2}\|(\bI \otimes \bX)-\bU^{t}_{k+1}-\dfrac{1}{n} \bar{\by}\bar{\by}^{H} \|_{F}^{2}\\
\end{array}
\end{equation}
\begin{equation}\label{updatel}
\begin{array}{ll}
\hat{\blam}^{t}_{k+1} = \hat{\blam}^{t}_{k} +\rho\left((\bI \otimes \bX^{t}_{k+1})-\bU^{t}_{k+1}-\dfrac{1}{n} \bar{\by}\bar{\by}^{H}\right)
\end{array}
\end{equation}
where $(\cdot)^{t}_{k}$ is used to denote the $k$-th iteration of the ADMM algorithm in correspondence of the $t$-th MM outer loop. Problems (\ref{updateX}) and (\ref{updateR}) have closed-form solutions which can be computed via the projection of appropriate matrices onto the respective feasible sets. Indeed, Problem (\ref{updateX}) can be equivalently cast as
\begin{equation}\label{newupdateX}
\begin{array}{ll}
\bU^{t}_{k+1}= \underset{{\bU}}{\rm arg\:min}\: \|\bU -   \bPsi\|_{F}^{2}\\
\quad \quad \quad \quad{\rm subject\ to}\quad {\bU}\succeq 0 \\
\end{array}
\end{equation}
$\bPsi=\left(\left(\bI \otimes \bX^{t}_{k}\right)+\dfrac{(\hat{\blam}^{t}_{k})}{\rho}- \dfrac{1}{n} \bar{\by}\bar{\by}^{H} \right)$. Hence, solving (\ref{updateX}) is tantamount to performing the orthogonal projection of the matrix  $\bPsi$ onto the set of the Positive SemiDefinite (PSD) matrices which can be computed as $\bU^{t}_{k+1}=\tilde{\bV}\rm{max}({\rm{diag}}(\tilde{\bU}),\bzero)\tilde{\bV}^{H}$, where ${\rm{diag}}(\tilde{\bU})$ and $\tilde{\bV}$ are the matrices containing the eigenvalues and the corresponding eigenvectors  of the matrix $\bPsi$, respectively. Similarly, the update step of $\bX$ in (\ref{updateR}) can be re-written as:
\begin{equation}\label{newupdateR}
\begin{array}{ll}
\bX^{t}_{k+1}= \underset{{\bX \in Toep }}{\rm  arg\:min}\:  \|\bX -   \bLam\|_{F}^{2}\\
\end{array}
\end{equation}
where $\bLam = \dfrac{1}{n}\displaystyle\sum_{i=1}^{n}\left(({{\bW_{i}}})^{t}_{k+1}-\dfrac{1}{\rho}{({\blam_{i}})^{t}_{k}} - \dfrac{1}{\rho n}({\bX}_{t})^{-1}\right)$, $(\bW_{i})^{t}_{k+1}$ and $({\blam_{i}})^{t}_{k}$ are the $i^{th}$ diagonal block (of size $m \times m$) of $\dfrac{1}{n} \bar{\by}\bar{\by}^{H}+\bU^{t}_{k+1}$ and $\hat{\blam}^{t}_{k}$, respectively. The solution to (\ref{newupdateR}) requires the projection of the matrix $\bLam$ onto the set of Toeplitz matrices which can be calculated by replacing the elements along the diagonals of the matrix $\bLam$ with the average along the respective diagonal elements.  Before concluding, it is worth pointing out that since the surrogate minimization problem in (\ref{n5}) is convex, it is guaranteed that ADMM converges to a supposed existing optimal unique solution to (\ref{n5}) (see Section $3.2$ in \cite{boydadmm}, \cite{admmnew}). The pseudocode of the proposed algorithm is shown below.
\begin{center}
\begin{tabular}{@{}p{15cm}}
\hline
\hline
\bf{Algorithm 1 : Pseudocode of ATOM 1 algorithm} \\
\hline
\hline
{\bf{Input}}: Samples $\{\by_{1}, \by_{2}, \cdots,\by_{n}\} $ and $\rho$ \\
{\bf{Initialize}}: Set ${t}, {k} = 0$. Initialize ${{\bX}_{0}}$, ${{\bX}^{t}_{0}}$ and $\hat{\blam}^{t}_{0}$.\\ 
{\bf{Repeat}}: Given $\bX_{t}$, perform the $t+1$-th step.\\
\hspace{9mm}$k \leftarrow 0$\\
\hspace{9mm}{\bf{Repeat}}: Given ${{\bX}}^{t}_{k}$ and $\hat{\blam}^{t}_{k}$,  perform the $k+1$-th step.\\
\hspace{9mm} 1) Obtain $\bU^{t}_{k+1}$ by projecting the matrix $\bPsi=\left(\left(\bI \otimes\bX^{t}_{k}\right)+\dfrac{{\hat{\blam}}^{t}_{k}}{\rho}- \dfrac{1}{n} \bar{\by}\bar{\by}^{H} \right)$ onto the \\
\hspace{9mm}set of PSD matrices.\\
\hspace{9mm}2) Obtain $\bX^{t}_{k+1}$ by projecting the matrix 
$\bLam= \dfrac{1}{n}\displaystyle\sum_{i=1}^{n}\left((\bW_{i})^{t}_{k+1}-\dfrac{1}{\rho}{(\blam_{i})^{t}_{k}} - \dfrac{1}{\rho n}({\bX}_{t})^{-1}\right)$ \\
\hspace{9mm} onto the set of Toeplitz matrices.\\
\hspace{9mm}3) $\hat{\blam}^{t}_{k+1} = \hat{\blam}^{t}_{k} +\rho\left((\bI \otimes \bX^{t}_{k+1})-\bU^{t}_{k+1}-\dfrac{1}{n} \bar{\by}\bar{\by}^{H}\right)$\\
\hspace{9mm} {\textbf{until convergence}}. \\
 Set ${\bX}_{t+1}= {\bX}^{t}_{k+1}$ \\
$t\leftarrow t+1$\\
{\textbf{until convergence}}\\
{\bf{Output}}: $\bR_{{\rm{\bf{ATOM1}}}}= \dfrac{\bX^{*}}{m}$, where $\bX^{*}$ denotes the value of $\bX_{t}$ at the completion of the outer loop. \\
\hline
\hline
\end{tabular}
\end{center}
From Algorithm 1 it can be seen that ATOM1 requires initialization of the matrices $\bX_{0}$, $\bX^{t}_{0}$ and $\hat{\blam}^{t}_{0}$. $\bX_{0}$ can be set using the initialization scheme discussed in \cite{melt} and, as $t=0$, $\bX^{t}_{0}$ can be set equal to $\bX_{0}$ while $\hat{\blam}^{t}_{0}$ can be constructed as $\hat{\blam}^{t}_{0} = \bV\bV^{T}$, where the elements of $\bV$ are drawn randomly from a uniform distribution over $[0,1]^{mn}$. For $t\geq1$, the matrices $\bX^{t}_{0}$ and $\hat{\blam}^{t}_{0}$ can be initialized with their last value after convergence at the previous ADMM iteration, respectively. Another input parameter required by ATOM1 is the penalty parameter $\rho$, introduced during the formation of the Augmented Lagrangian of  the ADMM algorithm. It is shown in \cite{boydadmm}, that the ADMM algorithm converges for any value of $\rho>0$. However, the numerical stability and the convergence rate of the algorithm depends on the choice of $\rho$. Simulation results have highlighted that for $\rho = m$, the ADMM algorithm is stable for different values of $n$ and $m$. Hence, unless otherwise stated, in all the numerical analysis the penalty parameter $\rho =m$ is used. 
\subsection{Computational complexity of ATOM1}\label{sec:3b}
ATOM1 is iterative in nature with two loops - the outer loop updates the Toeplitz matrix $\bX_{t}$ while the inner loop solves the surrogate minimization problem using ADMM.  Note that in the inner loop, one requires to compute the matrix $\dfrac{1}{n}\bar{\by}\bar{\by}^{H}$ - which is iteration independent and hence can be pre-computed. Moreover, this matrix can be evaluated efficiently leveraging an equivalent formulation of Problems (\ref{s1}) and (\ref{n5}). In this respect, first obtain the Cholesky decomposition of the matrix $\bR_{\textrm{SCM}}$, i.e., compute $\bR_{\textrm{SCM}} = \bar{\bR}\bar{\bR}^{H}$ (where $\bar{\bR}$ is a lower triangular matrix) with a complexity of  $\mathcal{O}(m^{3})$. Then, Problem (\ref{n5}) becomes
\begin{equation}
\begin{array}{ll}
\bX_{t+1}=\underset{{\bX}  \in Toep, \bX \succeq 0}{\rm arg\:min} \: \textrm{Tr}\left(({\bX}_{t})^{-1}{\bX}\right) \:\:
{\rm s.t.}\quad \dfrac{1}{n}\displaystyle\sum_{i=1}^{n}[\bar{\bR}]_{i}^{H}{\bX}^{-1}[\bar{\bR}]_{i} \leq 1
\end{array}
\end{equation}
Casting the above problem as an SDP it follows that
\begin{equation}\label{temp}
\begin{array}{ll}
\bX_{t+1}=\underset{{\bX}  \in Toep, \boldsymbol{\hat{X}}}{\rm arg\:min} \: \textrm{Tr}\left(({\bX}_{t})^{-1}{\bX}\right) \:\:
{\rm s.t.}\quad  \boldsymbol{\hat{X}} \geq  \bar{\bR}_{\textrm{SCM}}
\end{array}
\end{equation}
where $\bar{\bR}_{\textrm{SCM}}=\bar{\br}\bar{\br}^{H}$ and the vector $\bar{\br} \in \mathbb{C}^{m^{2}\times1}$ is obtained by vectorizing the matrix $\bar{\bR}$. Comparing the constraint of Problem (\ref{temp}) and (\ref{n5}), it can be observed that instead of computing the matrix $\dfrac{1}{n}\bar{\by}\bar{\by}^{H}$, one needs to evaluate $\bar{\bR}_{\textrm{SCM}}$. Exploiting the sparseness of the vector $\bar{\br}$, the matrix $\bar{\bR}_{\textrm{SCM}}$ can be determined with a cost
of $\mathcal{O}\left(\left(\dfrac{m(m+1)}{2}\right)^{2}\right)$ \cite{golub}. Let us now discuss the complexity related to the outer and inner loops of ATOM1. In this regard, note that the inner loop of ATOM1 requires the computation of the matrix $({\bX_{t}})^{-1}$ - which is outer loop iteration dependent. Therefore, this matrix can be computed once in the outer loop. Consequently, apart from the computations of the inner loop, an outer loop cycle requires only the estimation of the matrix $({\bX_{t}})^{-1}$. This in-turn can be efficiently computed using Gohberg–Semencul formula with a complexity $\mathcal{O}(m\:{\rm{log}}m)$ \cite{stoica}. The computational complexity of an inner loop cycle is related to the projection of $\bPsi$ onto the set of PSD matrices and projection of $\bLam$ onto the set of Toeplitz matrices. 
The cost of  projection of $\bLam$ onto the set of Toeplitz matrices is mainly dictated by the computation of average of the elements along the respective diagonals of $\bLam$. Hence, the cost of matrix projection onto the Toepltiz matrices set is $\mathcal{O}(m^{2})$. Next, the projection of $\bPsi$  onto the set of PSD matrices mainly involves the computation of the eigenvalues and eigenvectors of the matrix $\bPsi$ - whose corresponding complexity is $\mathcal{O}({m^{6}})$ \cite{golub}. Therefore, the computational complexity of ATOM1 is $\mathcal{O}(\eta{m^{6}})$ where $\eta$ is the total number of inner loop iterations required by the algorithm to converge.
A drawback of ATOM1 is that it has a restriction in handling additional constraints on the covariance matrix. This is because ATOM1 implements ADMM algorithm at every inner iteration which requires the optimization problem to exhibit the standard form \cite{boydadmm,admmsdp}
\begin{equation}\label{format}
\begin{array}{ll}
\underset{{\bZ}, \bE}{\rm minimize} \:& f(\bZ) + g(\bE) \\
{\rm subject\ to} & \:\: \bA_{1}\bZ+\bA_{2}\bE = \bC
\end{array}
\end{equation}
where $f(\bZ)$, $g(\bE)$ are convex functions and $\bA_{1}$, $\bA_{2}$, $\bC$ are matrices of appropriate dimensions, respectively. Therefore, to incorporate additional inequality constraints (such as an upper bound on the condition number of the matrix $\bZ$), one needs to replace the inequality constraint with an equality constraint. This can be done by introducing a slack variable to the existing optimization variables $\bZ$ and $\bE$. However, there is no convergence guarantee of ADMM when there are more than two optimization variables \cite{admmext}. This issue is addressed in the following by proposing another low complexity algorithm, referred to as ATOM2, to solve Problem (\ref{n5}).
\subsection{ATOM 2}\label{sec:3c}
To develop a computationally efficient and flexible estimation procedure capable of including additional constraints, a different surrogate function is constructed for the objective function in Problem ($P_{2}$). To this end, let us once again refer to Problem (\ref{n5}) cast in the equivalent (but computationally convenient) form
\begin{equation} \label{n5new}
\begin{array}{ll}
 \underset{{\bX\in Toep,{\boldsymbol{\hat{X}}}}}{\rm minimize} \: \textrm{Tr}\left(({\bX}_{t})^{-1}{\bX}\right)\\
{\rm subject\ to}\quad \boldsymbol{\hat{X}}\succeq \bar{\bR}_{\textrm{SCM}}
\end{array}
\end{equation}
and suitably constructing its surrogate function. In this respect, adding and subtracting $\textrm{Tr}({\bX}^{2})$ in the objective function yields
\begin{equation}
\begin{array}{ll}
\underset{{\bX}  \in Toep, \boldsymbol{\hat{X}}}{\rm minimize} \: \textrm{Tr}\left(({\bX}_{t})^{-1}{\bX}\right)+\textrm{Tr}({\bX}^{2})-\textrm{Tr}({\bX}^{2}) \\
{\rm subject\ to}\quad \boldsymbol{\hat{X}}\succeq \bar{\bR}_{\textrm{SCM}}
\end{array}
\end{equation}
Note that $-\textrm{Tr}({\bX}^{2})$ is a concave function of ${\bX}$ and hence given a feasible solution $\bX_{t}$, it is possible to construct the surrogate function $\tilde{g}(\bX|\bX_{t})$ for the above problem using Lemma \ref{lemma 0}
\begin{equation}\label{ns}
    \begin{array}{ll}
    \tilde{g}(\bX|\bX_{t}) =  \textrm{Tr}\left(({\bX}_{t})^{-1}{\bX}\right)+\textrm{Tr}({\bX}^{2})-2\textrm{Tr}({\bX}{{\bX}}_{t})
    \end{array}
\end{equation}
Note that $\tilde{g}(\bX|\bX_{t})$ is a surrogate to a surrogate function. Nonetheless, since $\tilde{g}(\bX|\bX_{t})$ is a tighter surrogate for $g(\bX|\bX_{t})$, it can be viewed as a direct surrogate for the objective function in Problem ($P_{2}$). Then, at any iteration $t$, given $\bX_{t}$, an optimal solution to the following surrogate minimization problem 
\begin{equation} \label{prefinal}
\begin{array}{ll}
\bX_{t+1}=\underset{{\bX}  \in Toep}{\rm arg\:min} \: \|\bX-\bB\|_{F}^{2}\\
\hspace{1.5cm}{\rm subject\ to}\quad \boldsymbol{\hat{X}} =\bI \otimes \bX  \succeq \bar{\bR}_{\textrm{SCM}}
\end{array}
\end{equation}
is required, where $\bB = {{\bX}}_{t} - 0.5({\bX}_{t})^{-1}$. 
An efficient way to solve Problem (\ref{prefinal}) is now devised. To this end, let us cast (\ref{prefinal}) in terms of a single block diagonal matrix $\bT$ as follows
\begin{equation} \label{final}
\begin{array}{ll}
\underset{{\bT}}{\rm minimize} \: \|\bT-\bI\otimes\bB\|_{F}^{2}\\
{\rm subject\ to}\:\: \bT \succeq \bar{\bR}_{\textrm{SCM}}\\
\hspace{17.5mm}{\bT\: \textrm{is a block diagonal matrix with the same diagonal Toeplitz blocks}}
\end{array}
\end{equation}
Problem (\ref{final}) seeks for the point $\bT$ belonging to the intersection of the two sets - the set of block diagonal matrices with the same diagonal Toeplitz blocks and the set defined by the Linear Matrix Inequality (LMI) $\bT \succeq \bar{\bR}_{\textrm{SCM}}$ - closest to $\bI\otimes\bB$. Since both the sets are convex, one can use the technique of alternating projection or Projection Onto the Convex Sets (POCS) to solve Problem (\ref{final}) \cite{ap1,pocsnew,pocs}. To describe this algorithm, let us denote by ${\mathcal{P}_{Toep}}(\bPsi)$ and ${\mathcal{P}_{LMI}}(\bPsi)$ the orthogonal projection of a matrix $\bPsi$ onto the set of block diagonal matrices with the same diagonal Toeplitz blocks and onto the set defined by $\bT \succeq \bar{\bR}_{\textrm{SCM}}$, respectively. Both ${\mathcal{P}_{Toep}}(\bPsi)$ and ${\mathcal{P}_{LMI}}(\bPsi)$ can be obtained as follows: 
 \begin{enumerate}
 \item{Calculation of ${\mathcal{P}_{Toep}}(\bPsi)$:\\} 
 Given a matrix $\bPsi$, the orthogonal projection of the matrix onto the set of block diagonal matrices with the same diagonal Toeplitz blocks can be obtained in two steps. First the averaging of the diagonal blocks of the matrix $\bPsi$ is performed. Then, the elements along each diagonal of the matrix obtained in the previous step are averaged.
 \item{Calculation of ${\mathcal{P}_{LMI}}(\bPsi)$:\\}
 Given a matrix $\bPsi$, its orthogonal projection onto the set defined by LMI $\bT \succeq \bar{\bR}_{\textrm{SCM}}$ can be obtained as follows. First, calculate the EigenValue Decomposition (EVD) of the matrix $\bPsi -\bar{\bR}_{\textrm{SCM}}$, i.e., obtain $[\bar{\bU}, \bar{\bV}] = {\rm{eig}}(\bPsi -\bar{\bR}_{\textrm{SCM}})$, where $\bar{\bU}$ and $\bar{\bV}$ are matrices containing the eigenvalues and eigenvectors  of the spectral decomposition, respectively. Then, the orthogonal projection ${\mathcal{P}_{LMI}}(\bPsi)$ is given by $\bar{\bV}{\rm{max}}(\bar{\bU},\bzero)\bar{\bV}^{H} + \bar{\bR}_{\textrm{SCM}}$.
 \end{enumerate}
According to POCS method, given an initial value $\bT_{0}=\bI\otimes\bB$, first compute $\bY_{k+1} ={\mathcal{P}_{Toep}}(\bT_{k})$ and then, using $\bY_{k+1}$, determine $\bT_{k+1}={\mathcal{P}_{LMI}}(\bY_{k+1})$, i.e., the starting point of the next iteration. Hence, POCS finds a sequence of iterates $\bT_{k}$ by alternatingly projecting between the two convex sets.  However, as reported in \cite{slowconvg}, POCS may suffer from slow convergence. A refinement of POCS is Dykstra's projection \cite{dykstra,dykstra2} which finds a point closest to $\bI\otimes\bB$ by adding correction vectors $\bP_{k}$  and $\bQ_{k}$ before every projection, which in-turn ensures convergence of sequence $\bT_{k+1}$ to the optimal solution ${\bT}^{*}$ \cite{dykstra}. The pseudocode of Dykstra's algorithm is shown in {\textbf{Algorithm 2}}.
 \begin{center}
\begin{tabular}{@{}p{16cm}}
\hline
\hline
\bf{Algorithm 2: Pseudocode of Dykstra's  algorithm} \\
\hline
\hline
{\bf{Input}}: $\bB = {{\bX}}_{t} - 0.5({\bX}_{t})^{-1}$ \\
{\bf{Initialize}}: Initialize $\bT_{0}= \bI \otimes \bB$, ${{\bP}_{0}}=\bzero$ and ${{\bQ}_{0}}=\bzero$\\ 
{\bf{Repeat}}: \\
\hspace{10mm}1) $\bY_{k} = {\mathcal{P}_{Toep}}(\bT_{k}+\bP_{k})$\\
\hspace{10mm}2) $\bP_{k+1} = \bT_{k}+\bP_{k}-\bY_{k}$\\
\hspace{10mm}3) $\bT_{k+1} = {\mathcal{P}_{LMI}}(\bY_{k}+\bQ_{k})$\\
\hspace{10mm}4) $\bQ_{k+1} = \bY_{k}+\bQ_{k}-\bT_{k+1}$\\
\hspace{10mm}$k\leftarrow k+1$\\
{\bf{until convergence}}\\
{\bf{Output}}: ${\bT}^{*} =\bT_{k+1}$.\\
\hline
\hline
\end{tabular}
\end{center}
 
Once the optimal solution $\bT^{*}$ is obtained via Dykstra's projection, the matrix ${\bX}_{t+1}$ can be constructed from its upper left block of size $m \times m$. This process is repeated until the whole algorithm, i.e., including the outer loop, converges. Finally, using the relation in (\ref{relation}), the optimal Toeplitz matrix $\bR^{*}$ of Problem ($P_{1}$) is attained. ATOM2 is summarized in {\textbf{Algorithm 3}}.
\begin{center}
\begin{tabular}{@{}p{16cm}}
\hline
\hline
\bf{Algorithm 3: Pseudocode of ATOM2} \\
\hline
\hline
{\bf{Input}}: Samples $\{\by_{1}, \by_{2}, \cdots,\by_{n}\} $ \\
{\bf{Initialize}}: Set \emph{t} = 0. Initialize $\bX_{0}$. \\ 
{\bf{Repeat}}: Given ${{\bX}_{t}}$ perform the $t+1$-th step.\\
\hspace{10mm}Run Dykstra's algorithm with $\bB = {{\bX}}_{t} - 0.5({\bX}_{t})^{-1}$ as input and obtain $\bX_{t+1}$ from \\ 
\hspace{10mm}the upper left block of $\bT^{*}$.\\
\hspace{10mm}$t\leftarrow t+1$\\
{\bf{until convergence}}\\
{\bf{Output}}: $\bR_{{\rm{\bf{ATOM2}}}}=\dfrac{\bX^{*}}{m}$, where $\bX^{*}$ denotes the value of $\bX_{t}$ at the completion of the outer loop. \\
\hline
\hline
\end{tabular}
\end{center}
ATOM2 requires the initialization of the matrix $\bX$. In this respect, a similar scheme as in ATOM1 is followed, i.e., at each outer iteration, the initial guess required to determine $\bX_{t+1}$ in the inner loop is set equal to $\bX_{t}$. Furthermore, at $t=0$, the initial value of the matrix $\bX_{0}$ is obtained according to the initialization scheme discussed in \cite{melt}. 
\subsection{Computational complexity of ATOM2}\label{sec:3d}
Similar to ATOM1, ATOM2 is also an iterative algorithm with outer and inner loops. The outer loop updates the Toeplitz matrix $\bX_{t}$ and the inner loop implements the Dykstra's algorithm - which requires the computation of the matrices $\bar{\bR}_{\textrm{SCM}}$ and $({\bX_{t}})^{-1}$. Note that, the former matrix is iteration independent and  therefore can be pre-computed with a cost of $\mathcal{O}\left(\left(\dfrac{m(m+1)}{2}\right)^{2}\right)$. On the other hand, the matrix $({\bX_{t}})^{-1}$ is outer loop iteration dependent and therefore can be computed once in the outer loop. Consequently, apart from the inner loop computations, the outer loop demands only the
computation of $({\bX_{t}})^{-1}$ -  which can be computed efficiently with complexity $\mathcal{O}(m\:{\rm{log}}m)$. Meanwhile, the computational load of the inner loop stems from the computation of EVD of the matrix $(\bY_{k} +\bQ_{k})$ plus a rank one matrix $\bar{\bR}_{\textrm{SCM}}$ - which has a complexity of about $\mathcal{O}(m^{6})$. However, as already pointed, at the expense of slightly slower convergence, one can also implement the inner loop of ATOM2 using  POCS algorithm.  In this case, it is demanded to compute the EVD of a structured matrix, i.e., a Toeplitz structured matrix $\bY_{k}$ plus a rank one matrix $\bar{\bR}_{\textrm{SCM}}$. This can be computed efficiently in $\mathcal{O}(m^{2}\:{\rm{log}}m +m^{2})$ \cite{secular,eg}. Therefore, for large values of covariance matrix dimension $m$ one can implement the inner loop of ATOM2 algorithm using POCS algorithm. Hence, the total computational cost of ATOM2 and ATOM2 using POCS (hereafter refereed to as ATOM2-POCS) is $\mathcal{O}(\eta(m^{6}))+\mathcal{O}(m^{4})$ and $\mathcal{O}(\eta(m^{2}\:{\rm{log}}(m)))+\mathcal{O}(m^{4})$, respectively and $\eta$ is used to represent the total number of inner loop iterations required by the algorithm to converge. In Table \ref{c1}, the computational complexity of ATOM1, ATOM2, and ATOM2-POCS is compared with that of the state-of-the-art iterative algorithms \cite{melt,em1}. Unlike the proposed algorithms, the state-of-the art methods are single loop iteration algorithms. Therefore, in the case of \cite{melt,em1} $\eta$ is used to represent the number of iterations required by the algorithm to converge. Inspection of Table \ref{c1} shows that ATOM1 and ATOM2 (whose inner loop is implemented using Dykstra's algorithm) has the highest complexity when compared to ATOM2-POCS, MELT and EM. As to ATOM2-POCS, the per cycle computational cost is only $m$ times larger than MELT and EM. Nevertheless, it is worth anticipating that this complexity increase is complemented by a superior performance in terms of covariance matrix MSE and achieved SINR.
\begin{table}[h]
\centering
\begin{tabular}{ | m{2cm} |m{3.1cm}| m{3.1cm}| m{2.1cm} | m{2.1cm}| } 
\hline
ATOM1& ATOM2&ATOM2-POCS& MELT \cite{melt}& EM \cite{em1}\\ 
\hline
$\mathcal{O}(\eta{m^{6}})$ & $\mathcal{O}(\eta{m^{6}})$& $\mathcal{O}(\eta(m^{2}\:{\rm{log}}(m)))+\mathcal{O}(m^{4})$ & $\mathcal{O}\left(\eta(m{\rm{log}}(m))\right)$ & $\mathcal{O}\left(\eta(m{\rm{log}}(m)\right))$\\ 
\hline
\end{tabular}
\caption{Comparison among computational complexity of ATOM1 and ATOM2 with other state-of-the-art iterative algorithms.}
\label{c1}
\end{table}
\subsection{Proof of convergence}\label{sec:3e}
In this subsection, the proof of convergence of ATOM1 and ATOM2 is established. In this regard, it is worth pointing out that both the algorithms differ in the way they construct and optimize the surrogate function for the Problem ($P_{2}$). Nonetheless, since ATOM1 and ATOM2 are based on the MM framework, the proof of convergence based on the following lemma will hold for both algorithms. \\
Before stating the lemma, let us first introduce the first-order optimality condition for minimizing a function over a convex constraint set.  A point $\bX$ is a stationary point of $f(\cdot)$ if $f'(\bX;\bD) \geq 0$ for all $\bD$ such that $\bX+\bD \in \mathcal{C}$, where $\mathcal{C}$ is the convex constraint set and  $f'(\bX;\bD)$ is the directional derivative of $f(\cdot)$ at point $\bX$ in direction $\bD$ and is defined as \cite{conv}
\begin{equation}
\begin{array}{ll}
f'(\bX;\bD) =\underset{\lambda \downarrow 0}{\lim} \: \textrm{inf}\:\dfrac{f(\bX+\lambda\bD) - f(\bX)}{\lambda}
\end{array}
\end{equation}
Based on the following lemma, both ATOM1 and ATOM2 are guaranteed to converge to a stationary point of Problem ($P_{2}$)
\begin{lemma}\label{lemma 3}
Denoting by $\{\bX_{t}\}$ the sequence generated by either ATOM1 or ATOM2, then the objective function of Problem ($P_{2}$) monotonically decreases along the iterations. Besides, any positive definite cluster point to $\{\bX_{t}\}$ is a stationary point to Problem ($P_{2}$).
\begin{IEEEproof}
See Appendix C for details.
\end{IEEEproof}
\end{lemma}
Before concluding this subsection, it is worth pointing out that $m \geq n/2$ is a sufficient condition to guarantee that $\{\bX_{t}\}$ is a bounded sequence whose cluster points are all positive definite matrices. Indeed, under such condition, the sequence $\{\bX_{t}\}$  belongs to a compact subset of positive definite matrices.
\subsection{\textbf{Extensions of ATOM2}}\label{sec:3f}
The extension of ATOM2 to handle additional constraints other than the Toeplitz structure in the covariance estimation process is now addressed. In particular, it is shown that ATOM2 can be generalized to account for the following scenarios: Banded Toeplitz matrices, Toeplitz-block-Toeplitz matrices, low rank matrices with Toeplitz structure plus a scalar matrix, and finally, matrices with Toeplitz structure satisfying a condition number constraint. However, as discussed at the end of Subsection \ref{sec:3b},  ATOM1 cannot be directly extended to tackle the general constraints as for instance an upper bound requirement to the condition number. \\ 
1. \textbf{MLE of low rank Toeplitz covariance term plus a scalar matrix} \\
In radar applications and in some array signal processing contexts \cite{eastr,melt,lr1}, the covariance matrix $\bR$ to be estimated often exhibits the following structure
\begin{equation}
\bR = \bR_{c}+\sigma\bI
\end{equation}
where $\bR_{c}$ denotes the covariance matrix of the interference (clutter or jammers) and $\sigma\bI$ is the covariance matrix of the thermal noise (where $\sigma>0$ represents the spectral density level of the white noise and is unknown). Also, $\bR_{c}$ usually exhibits a low rank Toeplitz structure with known rank $r \leq m$ \cite{updatetau}. As a result, the covariance estimation problem can be cast as
\begin{equation} \label{eq:12new}
\begin{array}{ll}
 \underset{\bR\in Toep, \bR_{c} \succeq 0, \sigma \geq 0}{\rm minimize} \:\dfrac{1}{n}\displaystyle\sum_{i=1}^{n}\by_{i}^{H}\bR^{-1}\by_{i} + \log|\bR|\\
\hspace{0.2cm}{\rm subject\ to}\quad\:\:\:{\bR}=\bR_{c} +\sigma\bI\\
\hspace{2.1cm} \quad{\rm Rank}({\bR_{c}}) \leq r
\end{array}
\end{equation}
Following a similar line of reasoning as that used in the development of ATOM2, the constrained ML estimate can be obtained by leveraging Lemma \ref{lemma 1} and solving the following equivalent problem
\begin{equation} 
\begin{array}{ll}
\underset{{\bX\in Toep,\bX_{c}\succeq 0,\boldsymbol{\hat{X}}},\sigma \geq 0}{\rm minimize} \: \log|\bX|\\
\hspace{0.5cm}{\rm subject\ to}\quad\:\:\boldsymbol{\hat{X}}\succeq \bar{\bR}_{\textrm{SCM}}\\
\quad \quad \quad \quad\quad\quad\quad{\bX}=\bX_{c}+\sigma\bI\\
\quad \quad \quad \quad\quad\quad\quad{\rm Rank}({\bX_{c}}) \leq r
\end{array}
\end{equation}
In this respect, the MM framework is exploited and, using Lemma \ref{lemma 0}, the following surrogate optimization problem is obtained
\begin{equation} \label{r1}
\begin{array}{ll}
\underset{{\bX \in Toep},\bX_{c}\succeq 0, \boldsymbol{\hat{X}},\sigma \geq 0}{\rm minimize} \: \|\bX-\bB\|_{F}^{2}\\
\hspace{0.5cm}{\rm subject\ to}\quad\:\:\boldsymbol{\hat{X}}\succeq \bar{\bR}_{\textrm{SCM}}\\
\quad \quad \quad \quad\quad\quad\quad{\bX}=\bX_{c}+\sigma\bI\\
\quad \quad \quad \quad\quad\quad\quad{\rm Rank}({\bX_{c}}) \leq r
\end{array}
\end{equation}
where $\bB = {\bX}_{t} - 0.5({\bX}_{t})^{-1}$.
It can be tackled using alternating projection among three subsets at each iteration: $\mathcal{S}_{1}=\left\{\bX: \boldsymbol{\hat{X}}\succeq \bar{\bR}_{\textrm{SCM}}\right\}$, $\mathcal{S}_{2}=\left\{\bX: {\bX}=\bX_{c}+\sigma\bI, \bX_{c}\succeq 0, {\rm Rank}({\bX_{c}}) \leq r, \sigma\geq 0\right\}$, and $\mathcal{S}_{3}=Toep$. Although there are no convergence guarantees for alternating projections in the presence of non-convex sets (the set of low rank matrices is non-convex set), this strategy usually proves effective, for instance see \cite{tropp} for the frame design problem. The projection of a given matrix onto $\mathcal{S}_{1}$ and $\mathcal{S}_{3}$ cane be performed as discussed in Section \ref{sec:3} B. Let $\hat{\bB}$ represent the matrix obtained after projecting the output of the previous algorithm iteration onto $\mathcal{S}_{1}$ (the algorithm is initialized with the input matrix $\bB$). Then, the projection of $\hat{\bB}$ onto $\mathcal{S}_{2}$ requires the solution of the following problem
\begin{equation}\label{p1-low rank}
    \begin{array}{ll}
    \underset{\bX,\bX_{c}\succeq 0,\sigma \geq 0}{\rm minimize}\:\:\: \|\bX-\hat{\bB}\|_{F}^{2}\\
\hspace{0.01cm}{\rm subject\ to}\:\:{\bX}=\bX_{c}+\sigma\bI\\
\quad \quad \quad \quad \quad {\rm Rank}({\bX_{c}}) \leq r
    \end{array}
\end{equation}
 Let ${\boldsymbol{\beta}} = [\beta_{1}, \beta_{2}, \cdots, \beta_{m}]^{T}$ and ${\boldsymbol{\gamma}}= [\gamma_{1}, \gamma_{2}, \cdots, \gamma_{m}]^{T}$ be the eigenvalues of the matrix $\hat{\bB}$ and $\bX$, sorted in decreasing order, respectively. Now, denoting by $\bU$ a unitary matrix such that $\hat{\bB}=\bU \rm{diag} (\boldsymbol{\beta})\bU^{H}$ and exploiting the unitary invariance of the objective and the constraint functions of (\ref{p1-low rank}), it follows that an optimal solution to (\ref{p1-low rank}) is $\bar{\bX}=\bU \rm{diag} ({\boldsymbol{\gamma}}^{*})\bU^{H}$ with $\boldsymbol{\gamma}^{*}=[\gamma_1^{*},\cdots,\gamma_r^{*},\sigma,...\sigma]^{T} \in \mathbb{R}^m$, the optimal solution to
\begin{equation}\label{p2-low rank}
    \begin{array}{ll}
     \underset{\gamma_{1}\geq\gamma_{r}\geq\sigma \geq 0}{\rm{minimize}}  \displaystyle\sum_{i=1}^{r}(\gamma_{i}- \beta_{i})^{2}+\displaystyle\sum_{i=r+1}^{m}(\sigma- \beta_{i})^{2}
    \end{array}
\end{equation}
The optimal solution to (\ref{p2-low rank}) is 
\begin{eqnarray}
\gamma^{*}_{i} &=& \beta_{i}, i =1,2 \cdots, r,\\
\sigma^{*} &=& \dfrac{1}{m-r} \displaystyle\sum_{i=r+1}^{m}\beta_{i}.
\end{eqnarray}
Hence $\bar{\bX} = \beta_{1}\bu_{1}\bu_{1}^{H} + \cdots \beta_{r}\bu_{r}\bu_{r}^{H} + \sigma^{*}(\bu_{r+1}\bu_{r+1}^{H} +\cdots+\bu_{m}\bu_{m}^{H})$.
\\
2. \textbf{MLE of banded Toeplitz covariance matrix}\\
The covariance matrix is constrained to exhibit a banded Toeplitz structure of bandwidth $b$ (see \cite{band, bt} for relevant applications). For instance, assuming a bandwidth $b=2$ and dimension $m=5$ the covariance matrix enjoys the following structure 
\[ \bR=
\begin{bmatrix}
    r_{1} & r_{2} & r_{3}& 0 & 0 \\
    r^{*}_{2} & r_{1}  & r_{2}  & r_{3} & 0\\ 
    r^{*}_{3} &  r^{*}_{2}  & r_{1}  & r_{2} &  r_{3}\\
    0 &  r^{*}_{3}  & r^{*}_{2}    & r_{1} &  r_{2}\\
    0 &  0& r^{*}_{3}  & r^{*}_{2}    & r_{1} 
\end{bmatrix}
\]
Then, the MLE problem for banded Toeplitz covariance matrix can be formulated as  
\begin{equation}
\begin{array}{ll}
 \underset{\bR  \in Band-Toep,\:\bR \succ 0}{\rm minimize} \:\dfrac{1}{n}\displaystyle\sum_{i=1}^{n}\by_{i}^{H}\bR^{-1}\by_{i} + \log|\bR|
\end{array}
\end{equation}
where $Band-Toep$ is used to denote the set of banded Toeplitz matrices. Exploiting again Lemma \ref{lemma 1}, the above problem can be cast in the following equivalent form
\begin{equation}\label{bttemp} 
\begin{array}{ll}
 \underset{{\bX\in Band-Toep, \boldsymbol{\hat{X}}}}{\rm minimize} \: \log|\bX|\\
\hspace{6mm}{\rm subject\ to}\quad \: \boldsymbol{\hat{X}}\succeq \bar{\bR}_{\textrm{SCM}}
\end{array}
\end{equation}
Hence (\ref{bttemp}) is handled via MM framework solving the following surrogate minimization problem
\begin{equation}\label{Bt}
\begin{array}{ll}
\underset{{\bT}}{\rm minimize} \: \|\bT-\bI\otimes\bB\|_{F}^{2}\\
{\rm subject\ to}\:\: \bT \succeq \bar{\bR}_{\textrm{SCM}}\\
\hspace{17.5mm}{\bT\: \textrm{is a block diagonal matrix with the same diagonal banded Toeplitz blocks}}
\end{array}
\end{equation}
The above problem involves two convex sets: the set defined by the LMI $\bT \succeq \bar{\bR}_{\textrm{SCM}}$ and the set of block diagonal matrices with each block being equal and having a banded Toeplitz structure with bandwidth $b$. Consequently, Dykstra's projection algorithm or POCS can be used to solve Problem (\ref{Bt}). The projection of a matrix onto the LMI set can be calculated as discussed earlier in Subsection \ref{sec:3c}. The projection of a matrix $\hat{\bPsi}$ onto the set of block banded Toeplitz matrices can be obtained in two steps. In the first step, averaging the diagonal blocks of the matrix $\hat{\bPsi}$ is performed. In the next step, the elements outside the bandwidth are replaced with zeros. In contrast, the elements within the bandwidth are replaced with the average of the respective diagonal elements of the matrix obtained in the first step \cite{projbt}.\\
3.  \textbf{MLE of Toeplitz-block-Toeplitz covariance matrix}\\
In space-time adaptive processing radar applications, the covariance matrix exhibits a Toeplitz-block-Toeplitz (TBT) structure. This refers to a block Toeplitz matrix with each block having a Toeplitz structure \cite{tbt, tbt2}. An example of a TBT-structured covariance matrix with $p$ blocks is shown below 
\[ \bR=
\begin{bmatrix}
    \bR_{1} & \bR_{2} & \dots  & \bR_{p} \\
    \bR^{*}_{2} & \bR_{1}  & \dots  & \bR_{2} \\
    \vdots & \ddots & \ddots & \vdots \\
    \bR^{*}_{p} & \dots & \bR^{*}_{2} & \bR_{1}
\end{bmatrix}
\]
The MLE problem of TBT covariance matrix is formulated as
\begin{equation}\label{TBTtemp}
\begin{array}{ll}
 \underset{\bR  \in TBT, \bR \succ 0}{\rm minimize} \:\dfrac{1}{n}\displaystyle\sum_{i=1}^{n}\by_{i}^{H}\bR^{-1}\by_{i} + \log|\bR|
\end{array}
\end{equation}
where the notation $TBT$ is used to indicate the set of Toeplitz-block-Toeplitz matrices. The minimizer of (\ref{TBTtemp}) is obtained by solving at any given step the following surrogate optimization problem
\begin{equation}\label{TBT}
\begin{array}{ll}
\underset{{\bT}}{\rm minimize} \: \|\bT-\bI\otimes\bB\|_{F}^{2}\\
{\rm subject\ to}\:\: \bT \succeq \bar{\bR}_{\textrm{SCM}}\\
\hspace{17.5mm}{\bT\: \textrm{is a block diagonal matrix with the same diagonal TBT blocks}}
\end{array}
\end{equation}
Problem (\ref{TBT}) exhibits two constraints - $1$) a LMI constraint and $2$) a structural constraint - where the optimization variable $\bT$ is confined to be a block diagonal matrix with each block being equal and having a TBT structure. Since both the constraints are convex, Dykstra's projection or POCS can be applied to solve Problem (\ref{TBT}). The projection of a matrix onto the LMI set can be calculated as discussed earlier in Section \ref{sec:3} B. The projection of a given matrix $\bar{\bPsi}$ onto the set of matrices having the TBT constraint can be obtained as follows. First, a matrix $\bar{\bPsi}_{\rm{Avg}}$ is obtained by averaging the diagonal blocks of the matrix $\bar{\bPsi}$. Next $p$ matrices are obtained by averaging the diagonal blocks of the matrix $\bar{\bPsi}_{\rm{Avg}}$ from the preceding step. Finally, each of the $p$ matrices are projected onto the Toeplitz set as described in Section \ref{sec:3} C \cite{projbt}. \\
4. \textbf{MLE of Toeplitz covariance matrix with condition number constraint} \\
In this extension, a condition number constraint $\kappa $  (assumed to be  known) in addition to the Toeplitz structural constraint on the covariance matrix (see for applications \cite{sol, cond2}) is considered. The MLE problem can be formulated as
\begin{equation}
\begin{array}{ll}
 \underset{\bR  \in Toep, \bR \succ 0}{\rm minimize} \:\dfrac{1}{n}\displaystyle\sum_{i=1}^{n}\by_{i}^{H}\bR^{-1}\by_{i} + \log|\bR|\\
 {\rm subject\ to}\:\:\: \dfrac{\lambda_{max}(\bR)}{\lambda_{min}(\bR)} \leq \kappa
\end{array}
\end{equation}
To get a minimizer of the above problem, the steps used in ATOM2 are borrowed to arrive at the following surrogate optimization problem
\begin{equation}\label{Tbtc}
\begin{array}{ll}
\underset{{\bT}}{\rm minimize} \: \|\bT-\bI\otimes\bB\|_{F}^{2}\\
{\rm subject\ to}\:\: \bT \succeq \bar{\bR}_{\textrm{SCM}}\\
\hspace{17.5mm}{\bT\: \textrm{is a block diagonal matrix with the same diagonal Toeplitz blocks}}\\ 
\hspace{17.5mm}\textrm{whose condition number is upper bounded by\:} \kappa
\end{array}
\end{equation}
The above problem includes three constraints: the LMI constraint, the structural constraint which confines the covariance matrix to follow a Toeplitz structure, and the condition number constraint.  All the three constraints are convex and therefore, Dykstra's projection or POCS can be applied to solve Problem (\ref{Tbtc}). The projection of a given matrix $\tilde{\bPsi}$ onto $\dfrac{\lambda_{max}(\tilde{\bPsi})}{\lambda_{min}(\tilde{\bPsi})} \leq \kappa$ is now discussed. Given a matrix $\tilde{\bPsi}$, let $\bV\bGam\bV^{H}$ denote its EVD with $\bGam = \rm{diag}([\gamma_{1},\gamma_{2}, \cdots, \gamma_{m}]^{T})$ where $\gamma_{1} \geq \gamma_{2} \geq \cdots \geq \gamma_{m}$ and the columns of $\bV$ contain the corresponding eigenvectors. Then, the projection of $\tilde{\bPsi}$ onto the condition number constraint is given by \cite{Augusto}: $\bV\bLam\bV^{H}$ where $\bLam = {\rm{diag}}(\blam(u^{*}))$ and $\blam(u)=[\lambda_{1}(u),\lambda_{2}(u), \cdots, \lambda_{n}(u)]^{T}$ with
\begin{equation}
    \lambda_{i}(u) = \min\left(\kappa u, \max\left(\gamma_{i},\max\left(0,u\right)\right)\right) \:\: i=1,2, \cdots, m
\end{equation}
and $u^{*}$ can be obtained by solving a convex optimization problem with a linear computational complexity, see \cite{sol} for more details. The projection onto the convex sets defined by the LMI and Toeplitz structure can be evidently handled as discussed in Subsection \ref{sec:3c}. 
\section{Cramer-Rao Lower bound calculation}\label{sec:crlb}
In this section, the Cramer-Rao Lower Bound (CRLB) is derived for three structured covariance matrices, namely, the Toeplitz, the banded Toeplitz, and the TBT structures. The CRLB provides a lower bound on the variance of any unbiased estimator \cite{kay}. To proceed further, let $\theta$ represent the real value vector parametrizing a given covariance matrix structure of interest. The specific definition of $\theta$ is provided in the next subsections for each case study. Then, the CRLB is the inverse of the Fisher Information matrix (FIM) whose $(i,k)^{th}$ element is
\begin{equation}
\begin{array}{ll}
[\bF]_{i,k} = \textbf{E}\left[\frac{\partial^{2} \log f(\by;\bR)}{\partial \theta_{i}\partial\theta_{k}}\right] 
\end{array}
\end{equation}
where $f(\by;\bR)$ is the likelihood function i.e., $f(\by; \bR) = \dfrac{1}{n}\displaystyle\sum_{i=1}^{n}\by_{i}^{H}\bR^{-1}\by_{i} + \log|\bR|$. The $(i,k)^{th}$ element of the FIM can be computed using the \emph{Slepian–Bangs
formula} \cite{stoica}
\begin{equation}\label{FIM}
[\bF]_{i,k} = n\textrm{Tr}\left(\bR^{-1}\frac{\partial\bR}{\partial \theta_{i}}\bR^{-1}\frac{\partial\bR}{\partial \theta_{k}}\right)
\end{equation}
In the following subsections, the lower bounds on $\bR(\bth)$ for some covariance structures considered in the manuscript are derived. 
\subsection{Toeplitz matrix}\label{sec:4a}
As the entries of the TSC matrix are completely characterized by its first row, i.e., $[r_{1}, r_{2},\cdots r_{m}]^{T}$, the covariance matrix $\bR \in \mathbb{H}^{m \times m}$ can be parameterized by $\bth = [r_{1}, \Re(r_{2}),\cdots\Re(r_{m}),\\ \Im(r_{2}),...,\Im(r_{m}) ]^{T} \in \mathbb{R}^ {2m-1}$ where $\Re(r_{i})$ and $\Im(r_{i})$ denotes the real and imaginary parts of $r_{i}$, respectively. Then, the covariance matrix $\bR$ can be expressed in terms of $\bth$ and basis matrices $\bB^{\rm{Toep}}_{g}$ (defined as in (\ref{basis})), $g=1,2,\cdots,m$ \cite{em2} 
\begin{equation}\label{toep}
    \begin{array}{ll}
    \bR = \displaystyle\sum_{g=1}^{m}\theta_{g}{\Re}(\bB^{\rm{Toep}}_{g}) + j \displaystyle\sum_{g=m+1}^{2m-1}\theta_{g}{\Im}(\bB^{\rm{Toep}}_{g-m+1})
    \end{array}
\end{equation}
The $(i,k)^{th}$ element of the matrix $\bB^{\rm{Toep}}_{g}$ is given as
\begin{equation}\label{basis}
\begin{array}{ll}
    [\bB^{\rm{Toep}}_{g}]_{i,k}= 
\begin{cases}
    1& i-k=g-1=0\\
    1+j & k-i=g-1\neq0\\
    1-j&i-k=g-1\neq 0\\
    0 &\rm{otherwise}
\end{cases}
\end{array}
\end{equation}
Using (\ref{toep}), $\frac{\partial\bR}{\partial \theta_{i}}$ can be  obtained as 
\[
    \frac{\partial\bR}{\partial \theta_{i}}= 
\begin{cases}
    \Re(\bB^{\rm{Toep}}_{i}) & 1\leq i \leq m\\
    j\Im(\bB^{\rm{Toep}}_{i-m+1})& m+1 \leq i \leq 2m-1
\end{cases}
\]
Substituting $\frac{\partial\bR}{\partial \theta_{i}}$ in (\ref{FIM}), yields the FIM for Toeplitz covariance matrix. Then, the CRLB on $r_{i}$ is 
\[
\begin{cases}
   [\bF^{-1}]_{i,i}& i=1\\
   [\bF^{-1}]_{i,i}+[\bF^{-1}]_{i+m-1, i+m-1}& i\neq1
\end{cases}
\]
\subsection{Banded Toepltiz matrix}\label{sec:4b}
In the case of banded Toeplitz matrix with bandwidth $b$, the first row of the covariance matrix $\bR  \in \mathbb{H}^{m \times m}$ has only $b+1$ non-zero terms. Therefore, $\bR$ can be parameterized via $\bth = [r_{1}, \Re(r_{2}),\cdots\Re(r_{b+1})\\,\Im(r_{2}),...,\Im(r_{b+1}) ]^{T} \in \mathbb{R}^ {2b+1}$. Besides $\bR$ can be expressed in terms of basis matrices $\bB^{\rm{Toep}}_{g}$ and real coefficients $\bth$ 
\begin{equation}
    \begin{array}{ll}
    \bR = \displaystyle\sum_{g=1}^{b+1}\theta_{g}{\Re}(\bB^{\rm{Toep}}_{g}) + j \displaystyle\sum_{g=b+2}^{2b+1}\theta_{g}{\Im}(\bB^{Toep}_{g-b})
    \end{array}
\end{equation}
and consequently 
\[
    \frac{\partial\bR}{\partial \theta_{i}}= 
\begin{cases}
    {\Re}(\bB^{\rm{Toep}}_{i}) & 1\leq i \leq b+1\\
    j{\Im}(\bB^{\rm{Toep}}_{i-b})& b+2\leq i \leq 2b+1
\end{cases}
\]
Substituting $\frac{\partial\bR}{\partial \theta_{i}}$ in (\ref{FIM}), yields the FIM for banded Toeplitz covariance matrix and the CRLB on $r_{i}$ can be obtained as
\[
\begin{cases}
   [\bF^{-1}]_{i,i}& i=1\\
   [\bF^{-1}]_{i,i}+[\bF^{-1}]_{i+b, i+b}& i\neq1
\end{cases}
\]
\subsection{Toeplitz-block-Toeplitz matrix}\label{sec:4c}
A TBT matrix with $p$ blocks of size $l$ can be specified by its first row i.e., $[r_{1}, {r_{2}}, \cdots, {r_{l}}, \cdots , r_{(p-1)(2l-1)+1},\\ {r_{(p-1)(2l-1)+2}}, \cdots , {r_{(p-1)(2l-1)+l}}]^{T}$. As a consequence, a TBT matrix can be expressed as follows
\begin{equation}
  \begin{array}{ll}
  \bR^{\rm{TBT}} = \bC_{0}\otimes\bR_{0} +\displaystyle\sum_{w=1}^{{p-1}} \left(\left(\bC_{w}\otimes\bR_{w}\right) +\left(\bC_{w}^{T}\otimes\bR^{*}_{w}\right)\right)
   \end{array}
\end{equation}
where $(i,k)^{th}$ element of the matrix $\bC_{w} \in \mathbb{R}^{l\times l}$ is given by
\[
    [\bC_{w}]_{i,k}= 
\begin{cases}
    1& i-k=w\\
    0 & \rm{otherwise}
\end{cases}
\]
and $\bR_{w}$ is
\begin{equation}
    \begin{array}{ll}
    \bR_{w} = \displaystyle\sum_{g=1}^{l}\theta_{(g +qw)}{\Re}(\bB^{\rm{Toep}}_{g}) + j \displaystyle\sum_{g={l+1}}^{2l-1}\theta_{(g+qw)}{\Im}(\bB^{Toep}_{g-l+1})
    \end{array}
\end{equation}
where $\bth = [r_{1}, \Re({r_{2}}), \cdots, \Re({r_{l}}), \Im({r_{2}}), \cdots, \Im({r_{l}}), \cdots, r_{(p-1)(2l-1)+1}, \Re({r_{(p-1)(2l-1)+2}}), \cdots , \Re({r_{(p-1)(2l-1)+l}}),\\ \Im({r_{(p-1)(2l-1)+2}}), \cdots, \Im({r_{(p-1)(2l-1)+l}})]^{T} \in \mathbb{R}^{(2l-1)p}$ and $\bB^{\rm{Toep}}_{g}$ of size $\l \times l$ is defined in (\ref{basis}). Let $\theta_{i}$ belong to the $z^{th}$ block and $q = 2l-1$,  then $\frac{\partial\bR}{\partial \theta_{i}}$ is given by 
\[
    \frac{\partial\bR}{\partial \theta_{i}}= 
\begin{cases}
    \bC_{0} \otimes \Re(\bB^{\rm{Toep}}_{i}) & 1\leq i\leq{l}, z=0\\
    j\bC_{0} \otimes \Im(\bB^{\rm{Toep}}_{i-l+1})& {l}+1\leq i \leq q, z=0\\
    \bC_{z} \otimes\Re(\bB^{\rm{Toep}}_{i-qz}) + \bC_{z}^{T} \otimes{\Re}(\bB^{\rm{Toep}}_{i-qz})  & 1+qz\leq i \leq {l}+qz, z \neq 0 \\
     j\bC_{z} \otimes{\Im}(\bB^{\rm{Toep}}_{i-qz-l+1}) -j \bC_{z}^{T} \otimes{\Im}({\bB^{*}}^{{\rm{Toep}}}_{i-qz-l+1})  & {l}+1+qz \leq i \leq q+qz, z\neq 0\\
\end{cases}
\]
Substituting $\frac{\partial\bR}{\partial \theta_{i}}$ in (\ref{FIM}), the FIM for TBT covariance matrix is obtained. Then, the CRLB on $r_{i}$ is 
\[
\begin{cases}
   [\bF^{-1}]_{2i-(z+1),2i-(z+1)}& i=1, l+1, \cdots, (p-1)l +1\\
  [\bF^{-1}]_{i+2z,i+2z}+[\bF^{-1}]_{i+l-1+2z, i+l-1+2z} & i\neq1, l+1, \cdots, (p-1)l +1\\
  
\end{cases}
\]
\section{Numerical Simulations}\label{sec:4}
In this section, the performance of the proposed Toeplitz covariance matrix estimators ATOM1 and ATOM2 is analyzed in comparison with some state-of-the-art algorithms via numerical simulations using the MSE metric. In particular, ATOM1 and ATOM2 are compared with the EM-based algorithm \cite{em1}, MELT \cite{melt}, and the SCM estimator. Furthermore, ATOM2 is also compared with techniques for estimating the covariance matrix with banded Toeplitz, TBT, and condition number constraints. Finally, the performance of the estimators is evaluated in terms of maximum achievable SINR in a typical radar signal processing scenario. All the algorithms are implemented in MATLAB using a PC with $2.40$ GHz processor and $16$ GB RAM.
\subsection{Assessment of iterative algorithms convergence for on-grid and off-grid frequencies}\label{sec:5a}
\begin{figure}[h]
\centering
\begin{subfigure}[c]{0.45\textwidth}
\centering
\includegraphics[height=2.3in,width=3.1in]{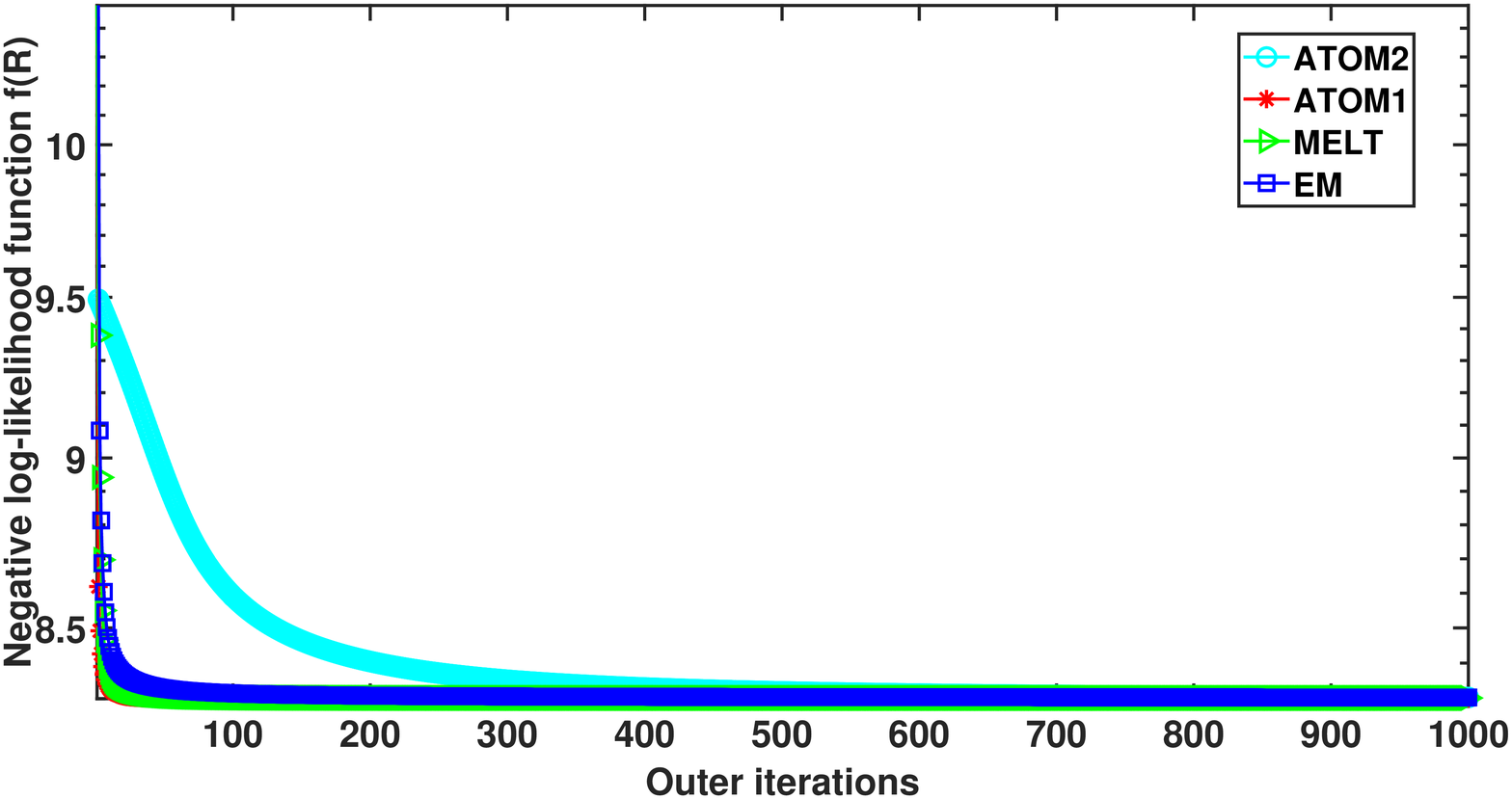}
\caption{}
\end{subfigure}
\hspace{6mm}
\begin{subfigure}[c]{0.45\textwidth}
\centering 
\includegraphics[height=2.3in,width=3.1in]{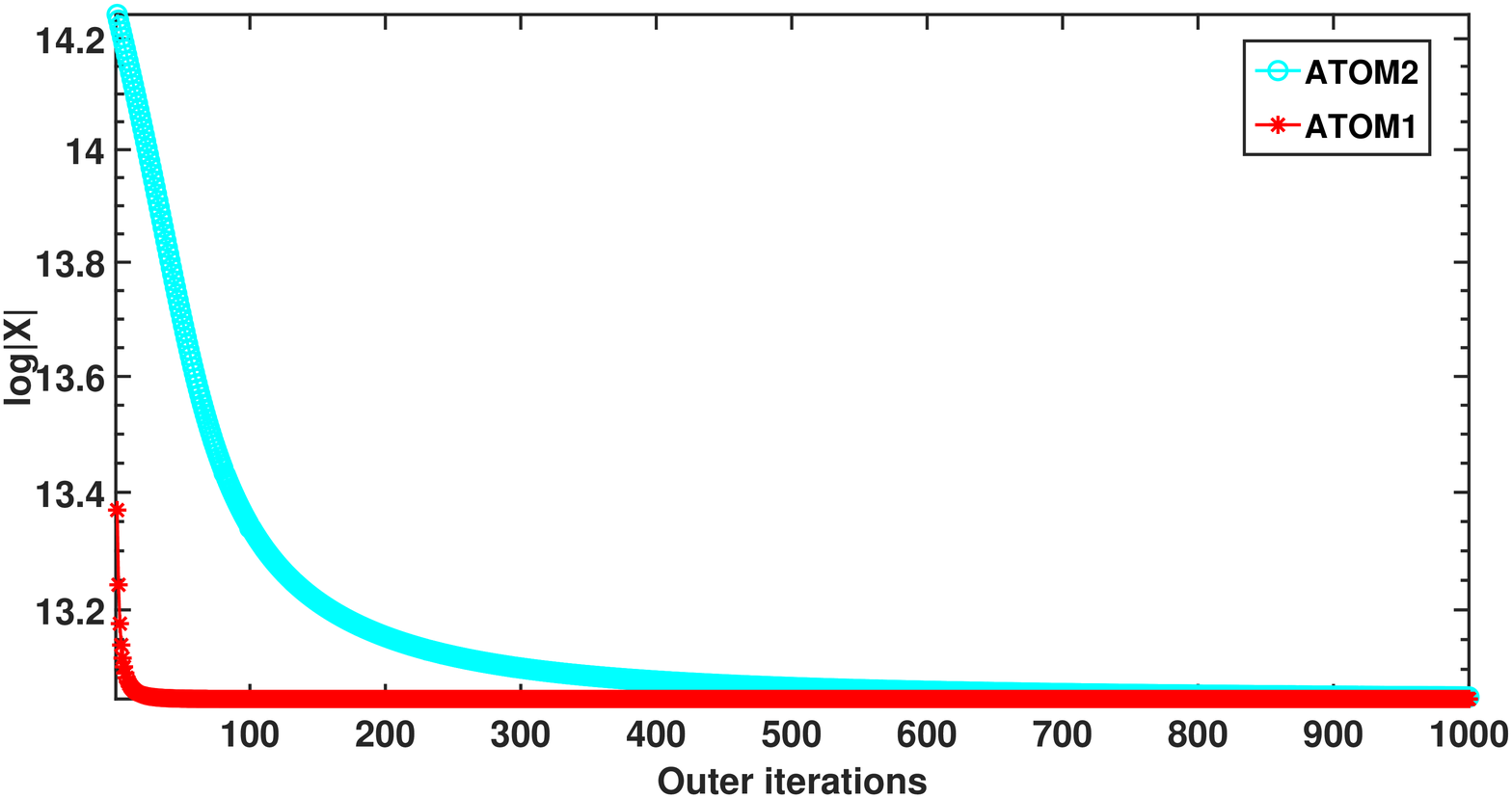}
\caption{}
\end{subfigure}
\caption{On-grid frequencies scenario for $m=6$ and $n=460$. a) Negative log-likelihood $f(\bR)$ vs. outer iterations; b) $\log|\bX|$ vs. outer iterations}
\label{obj}
\end{figure}

\begin{figure}[h]
\centering
\begin{subfigure}[c]{0.45\textwidth}
\centering
\includegraphics[height=2.3in,width=3.1in]{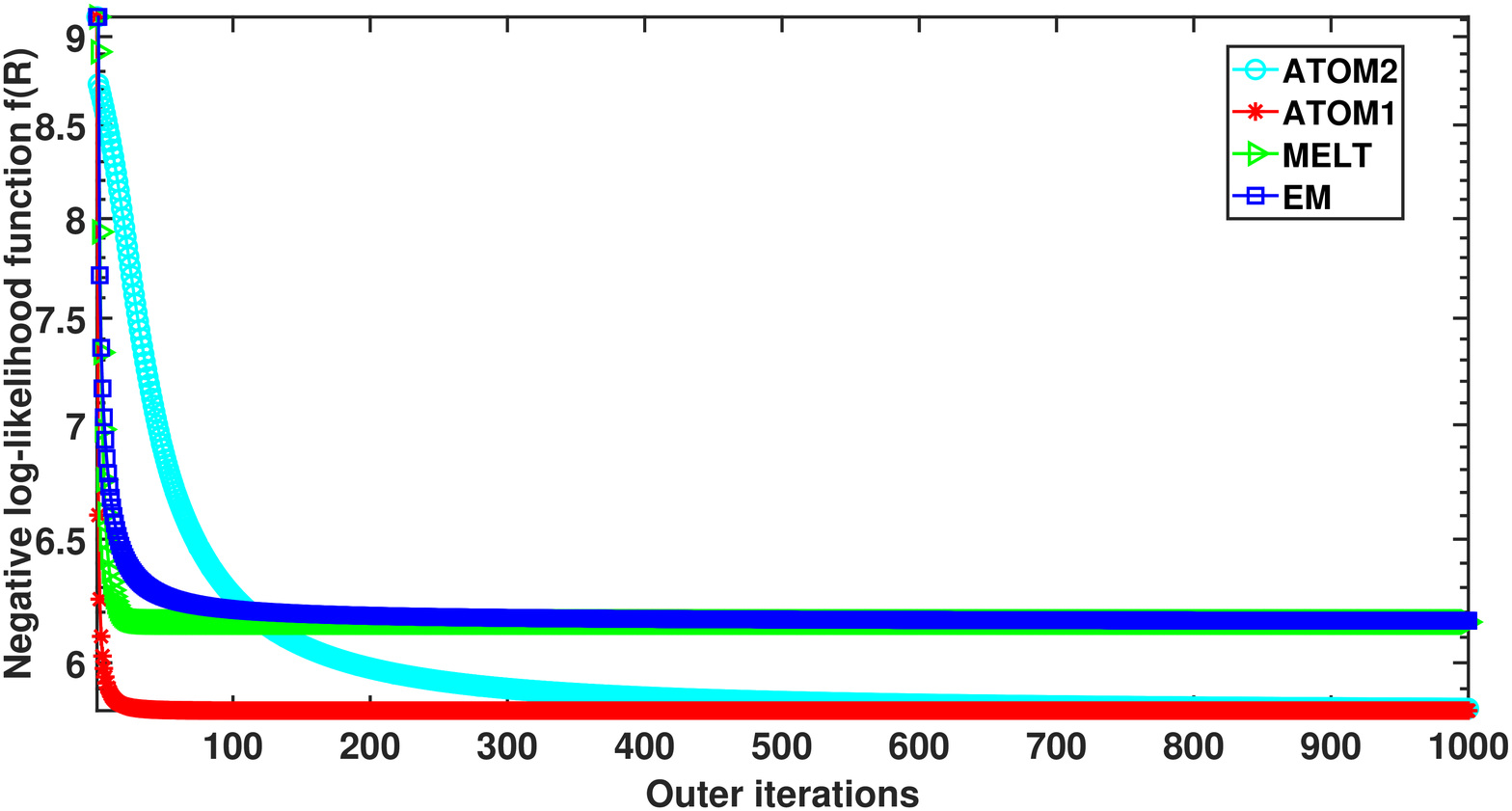}
\caption{}
\end{subfigure}
\hspace{6mm}
\begin{subfigure}[c]{0.45\textwidth}
\centering 
\includegraphics[height=2.3in,width=3.1in]{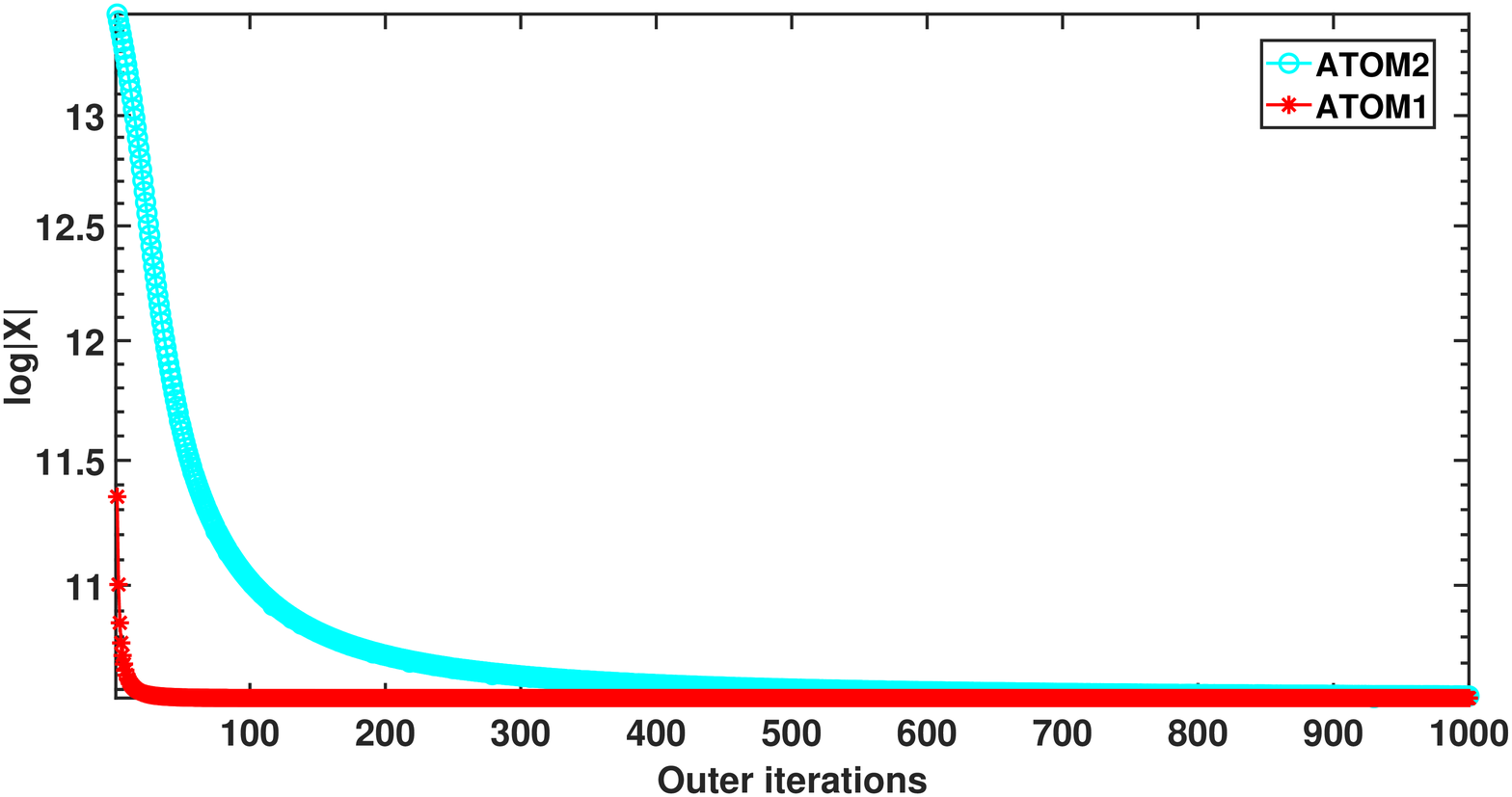}
\caption{}
\end{subfigure}
\caption{Off-grid frequencies scenario for $m=6$ and $n=460$. a) Negative log-likelihood $f(\bR)$ vs. outer iterations; b) $\log|\bX|$ vs. outer iterations}
\label{P2}
\end{figure}

In this simulation, the convergence of the ATOM1 and ATOM2 (whose inner loop was implemented via Dykstra's algorithm) is assessed in comparison with MELT and EM algorithms. To this end, the data $\by_{k}$ are generated according to the following model
\begin{equation}\label{data}
    \begin{array}{ll}
    \by_{k}= \sqrt{\bR}\bn_{k}, k=1,2, \cdots, n
    \end{array}
\end{equation}
where $\bn_{k}$'s are drawn randomly from a zero-mean circularly symmetric Gaussian distribution with independent and identically distributed entries with unit  mean square value and the size of the training data is $n=460$. The convergence of the iterative algorithms for the two different experimental setups are now evaluated. In the first setup, the true underlying Toeplitz covariance matrix $\bR$ of dimension $m=6$ is constructed by choosing the $2$-{nd}, $3$-rd, $5$-th, $7$-th, $8$-th and the $11$-th columns of the DFT matrix with $L=2m-1$ in (\ref{ce}), whose corresponding frequencies and amplitudes are: $[0.5712, 1.1424, 2.2848, 3.4272, 3.9984, 5.7120]$ rad/sec and $[3, 6, 4, 1,  7, 5]$, respectively. Fig. \ref{obj}.a shows the Objective Value (OV) of Problem $(P_{1})$, i.e., the negative log likelihood, versus the number of iterations for the first experimental setup. It can be seen that all the algorithms numerically decrease the underlying cost function and converge to the same OV. For reference, the OV of Problem $(P_{2})$, i.e., $\log|\bX|$ versus the number of iterations of ATOM1 and ATOM2 is also shown in Fig. \ref{obj}.b. This figure indicates that the proposed algorithms monotonically decrease the OV - which is expected since these algorithms optimize Problem $(P_{2})$ using the MM framework. In the second experimental setup, the true underlying Toeplitz covariance matrix is constructed such that one of the frequencies does not lie on the Fourier grid. The experimental setting is same as in case study 1, with the exception that the Fourier frequency $2.2848$ rad/sec is replaced with $2.5$ rad/sec. Fig. \ref{P2}.a shows the OV of Problem $(P_{1})$ versus the number of iterations for the second case study.  Fig. \ref{P2}.b indicates the corresponding OV of Problem $(P_{2})$ versus the number of iterations of the proposed methods. From Fig. \ref{P2}.a, it can be seen that while MELT and EM converges to a value of $6.16$, ATOM1 and ATOM2 converges to $5.81$. Therefore, when one of the frequencies does not lie on the Fourier grid, the state-of-the-art iterative algorithms converge to a larger value of the negative log-likelihood as compared to the proposed algorithms. This is because unlike the counterparts, the proposed algorithms estimate the Toeplitz covariance matrix without reparametrizing it via the CE technique and thus covers the whole set of Toeplitz covariance matrices. \\
Next, the computational time of the proposed techniques is compared with the other state-of-the-art iterative algorithms. The data samples $\by_{k}$ are generated using (\ref{data}) and $n=50$ samples. The Toeplitz covariance matrix $\bR$ is generated according to the model in (\ref{ce}) with $L=2m-1$ for four different values of $m$ - $4$, $8$, $16$ and $32$. The iterative algorithms have been run until the following condition is met 
\begin{equation}
\dfrac{\|\bR_{t}-\bR_{t+1}\|_{F}}{\|\bR_{t}\|_{F}} \leq 10^{-4}
\end{equation}
or until the maximum number of iterations (set equal to $1000$) is reached. Table. \ref{vary m} compares the average computational time (averaged over $50$ Monte-Carlo trials) of the different algorithms. The results show that ATOM2 has a smaller execution time than ATOM1. This is because the inner loop of ATOM1 (implemented via ADMM) requires an higher number of iterations and hence a longer run time to converge than ATOM2 inner loop involved in Dykstra's algorithm. From the table, it can also be seen that MELT has the least computational time. Nevertheless, although the proposed algorithms have a longer computational time than the counterparts, the obtained covariance matrix,  performs, in general, better in terms of MSE than the solution provided by MELT and EM. 
\begin{table}[h]
\centering
\caption {Comparison of the average run time (in seconds) of the iterative algorithms}
\label{vary m}
\begin{tabular}{|p{2.8cm}|p{1.8cm}|p{1.8cm}|p{1.8cm}|p{1.8cm}|}
\hline
Covariance matrix dimension $m$& ATOM1& ATOM2  &MELT\cite{melt}&EM\cite{em1} \\
 \hline
4&0.41&0.01&0.002&0.008\\
8&2.62&0.054&0.006&0.025\\
16&33.16&0.506&0.031&0.045\\
32&154.56&1.348&0.1043&0.2208\\
 \hline
\end{tabular} 
\end{table}
\subsection{MSE vs $n$ for Toeplitz covariance matrix}\label{sec:5b}
\begin{figure}[h]
\centering
\begin{subfigure}[c]{0.45\textwidth}
\centering
\includegraphics[height=2.5in,width=3.0in]{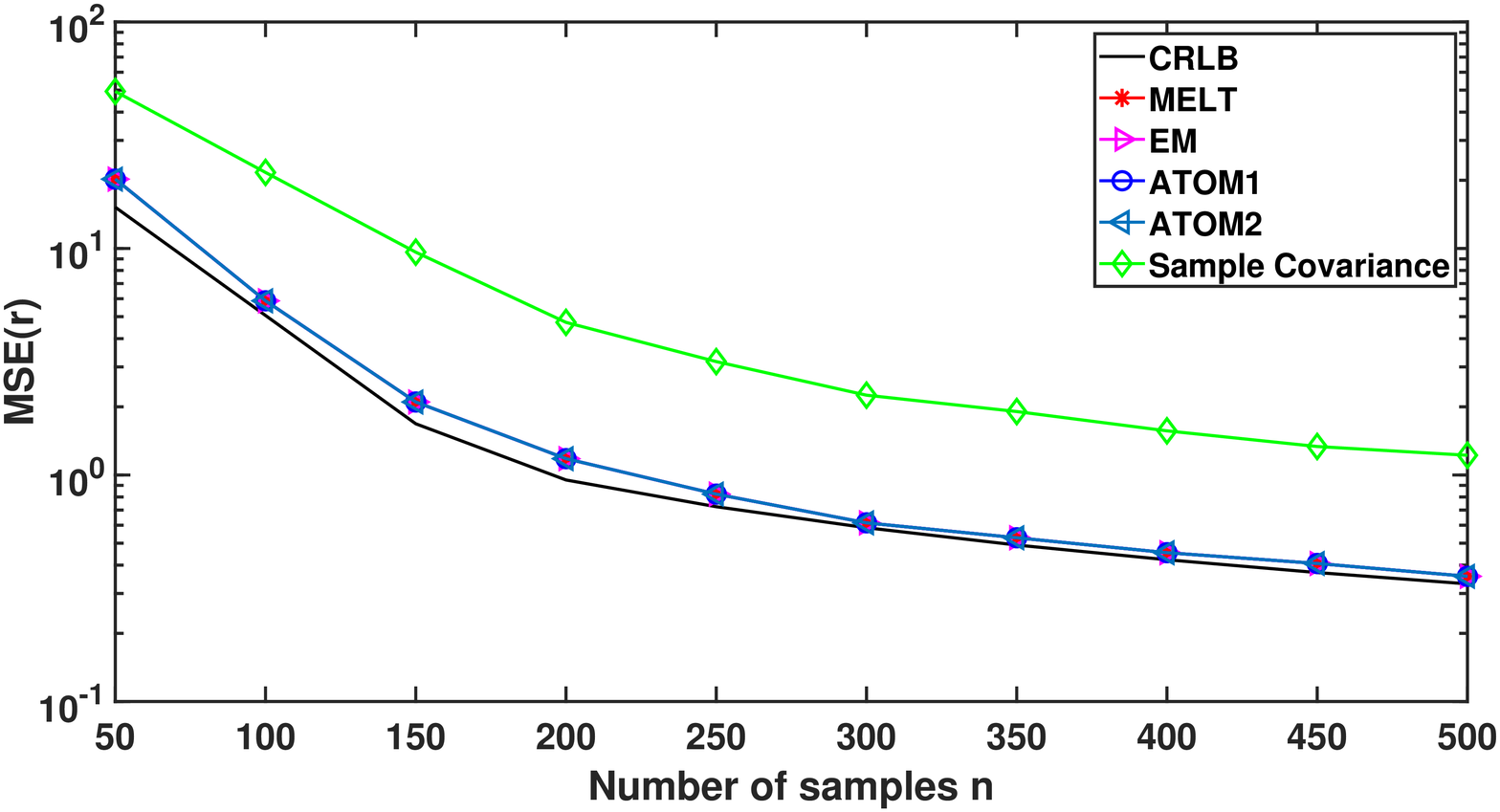}
\caption{}
\end{subfigure}
\hspace{5mm}
\begin{subfigure}[c]{0.45\textwidth}
\centering
\includegraphics[height=2.5in,width=3.0in]{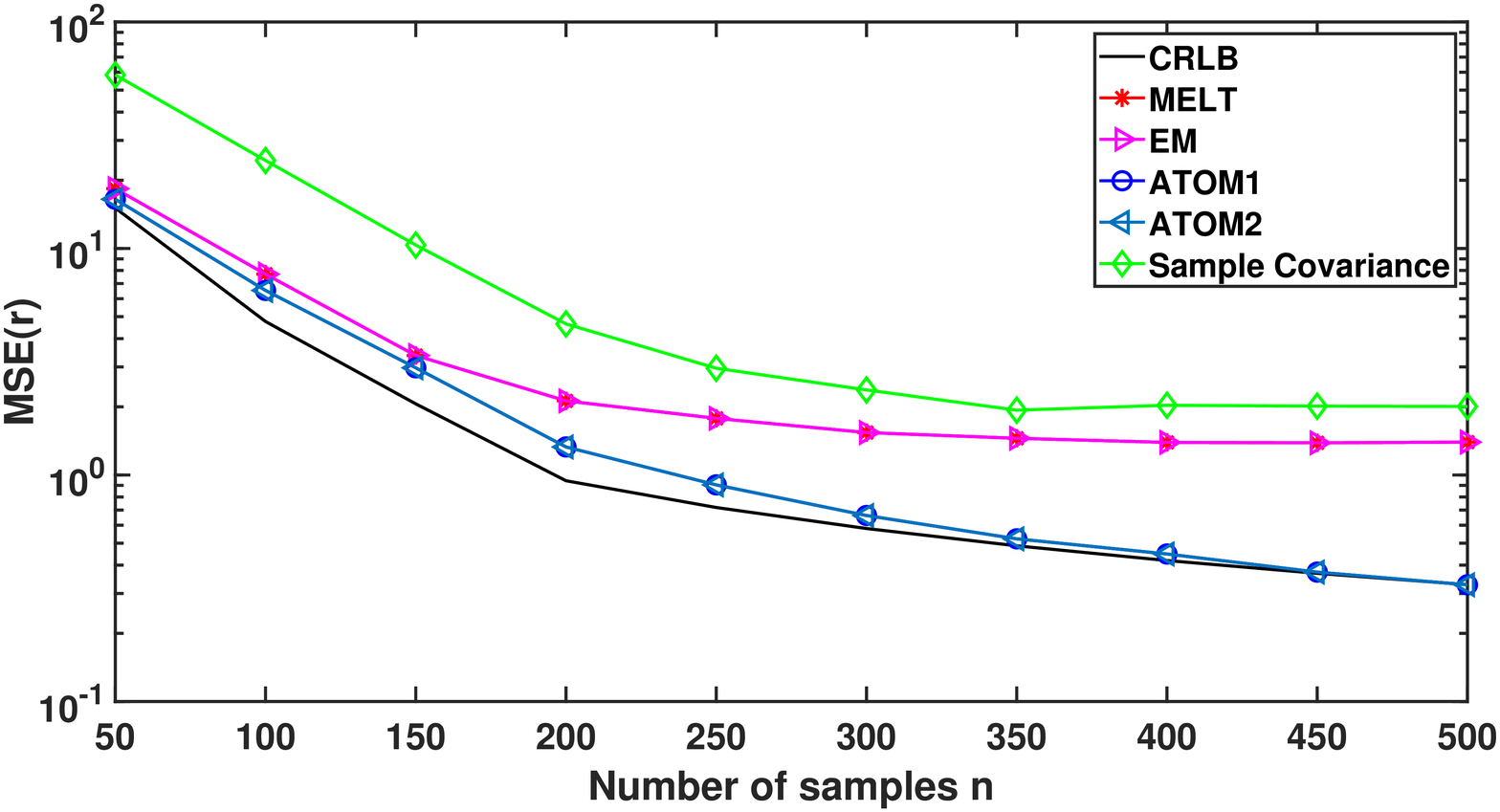}
\caption{}
\end{subfigure}
\caption{MSE vs. number of samples $n$ for Toeplitz covariance matrix. a) on-grid frequencies; b) off-grid frequencies}
\label{exp1}
\end{figure}

Data $\by_{k} \in \mathbb{C}^{15}$ are generated in the same way as in the previous subsection, modeling $\bR$ as in (1) with specific frequencies and amplitudes. The number of samples $n$ ranges between $50$ and $500$ in steps of $50$. As in the previous subsection, two different experiments are conducted assuming that the true Toeplitz covariance matrix is generated using on-grid and off-grid frequencies, respectively.  The algorithms are compared using MSE metric
\begin{equation}\label{mse}
\textrm{MSE}(\br) =      \textbf{E}\left[\dfrac{1}{m}\displaystyle\sum_{i=1}^{m}\left|r_{i}- \hat{r}_{i}\right|^2\right] 
\end{equation}
where $\br$ and $\hat{\br}$ indicate the first row of the true covariance matrix $\bR$ and its estimate $\hat{\bR}$, respectively. The number of Monte-Carlo trials used in this experiment to evaluate numerically (\ref{mse}) is equal to $100$. The sum of CRLB on each $r_{i}$ derived in Section \ref{sec:crlb} is used as benchmark.  
Fig. \ref{exp1}.a shows that in the first experiment\footnote{The frequencies used in the first experiment are: $[0.2167, 0.6500, 1.0833, 1.3, 1.5166, 1.9500, 2.3833, 2.8166, 3.2499, 3.6832, \\4.1166, 4.5499, 4.9832, 5.4165, 5.8499
]$ rad/sec. Their corresponding amplitudes increase linearly from $1$ to $15$ with a unit step.}, ATOM 1, ATOM 2, EM and MELT achieve a similar performance and reach the least MSE as compared to SCM. Fig. \ref{exp1}.b highlight that  in the second experiment (where the frequency $0.2167$ rad/sec is replaced with off-grid frequency $0.5$ rad/sec), ATOM 1 and ATOM 2 perform better than MELT, EM and SCM and, unlike the counterparts, achieve the CRLB when $n$ is sufficiently large. Furthermore, MELT and EM exhibit similar MSE's. Hence, unlike MELT and EM, the performance of ATOM 1 and ATOM 2 do not depend on the Fourier gridding. 
\vspace{-5mm}
\subsection{MSE vs $n$ for banded Toeplitz covariance matrix}\label{sec:5c}
\begin{figure}[!h]
\centering
\includegraphics[height=2.5in,width=3.0in]{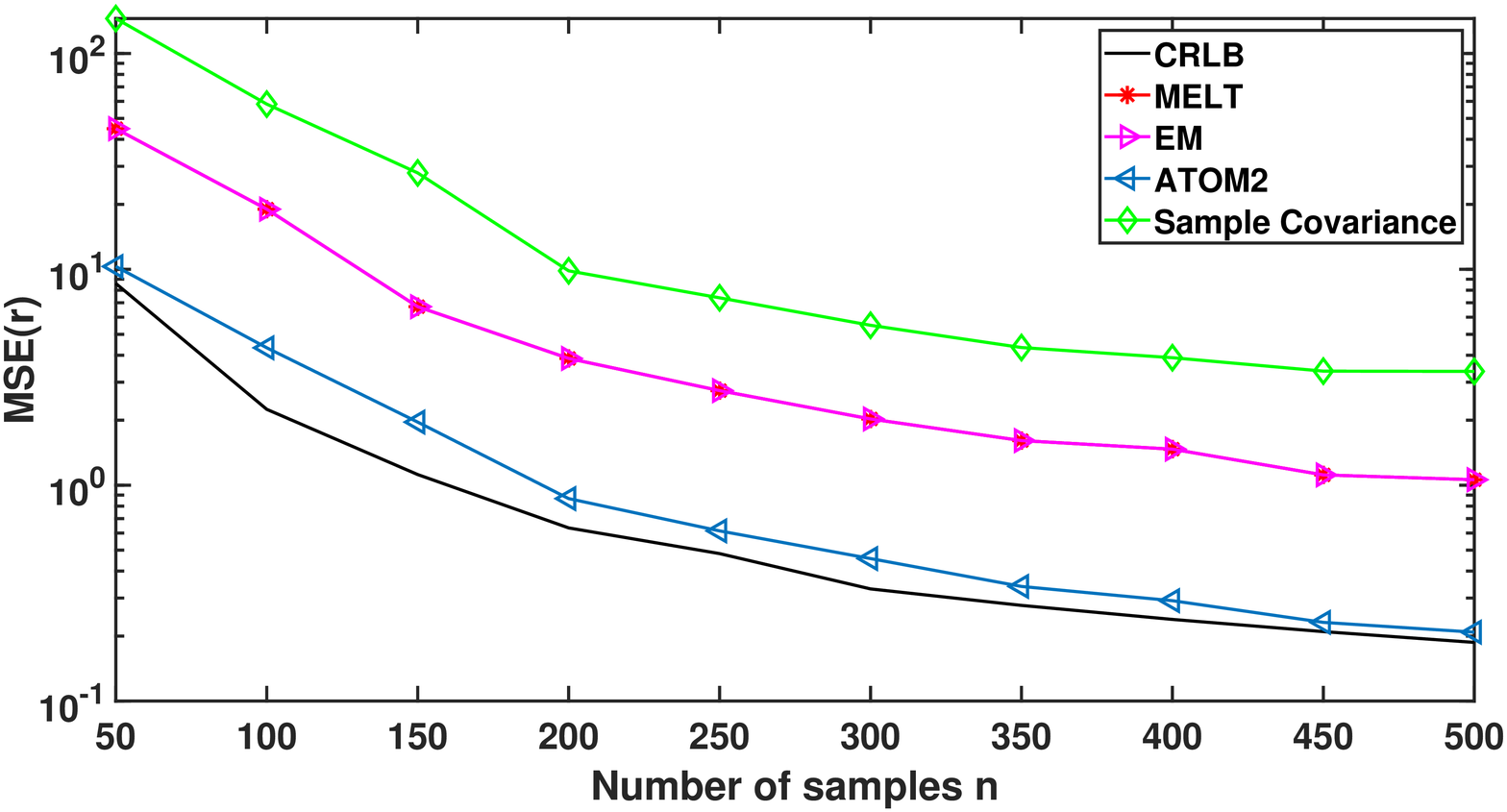}
\caption{MSE($\br$) vs. number of samples $n$ for banded Toeplitz covariance matrix}
\label{banded}
\end{figure}

In this subsection, the performance of ATOM2 is analyzed with the state-of-the-art algorithms when the covariance matrix is constrained to be a banded Toeplitz matrix. The ground truth banded Toeplitz matrix of dimension $m=15$ and bandwidth $b=6$ is constructed by alternately projecting a  random Hermitian matrix onto the set of banded Toeplitz matrices and the set of PSD matrices. Hence, the data $\by_{k}$ are generated using (\ref{data}). The number of samples $n$ ranges between $50$ and $500$ in steps of $50$. The number of Monte-Carlo experiments is the same as in the previous simulation 
and different algorithms are compared with in terms of MSE using the CRLB as benchmark. For each Monte-Carlo experiment the same ground truth covariance matrix was employed. Fig. \ref{banded} highlights that the proposed algorithm provide a lower MSE than the state-of-the-art algorithms.  This is because unlike the counterparts, the new method is developed without the CE. Also, as in the previous subsection, from Fig. \ref{banded} it can be observed that MELT and EM achieve a similar performance. 

\subsection{MSE vs $n$ for TBT covariance matrix}\label{sec:5d}
\begin{figure}[!h]
\centering
\includegraphics[height=2.5in,width=3.0in]{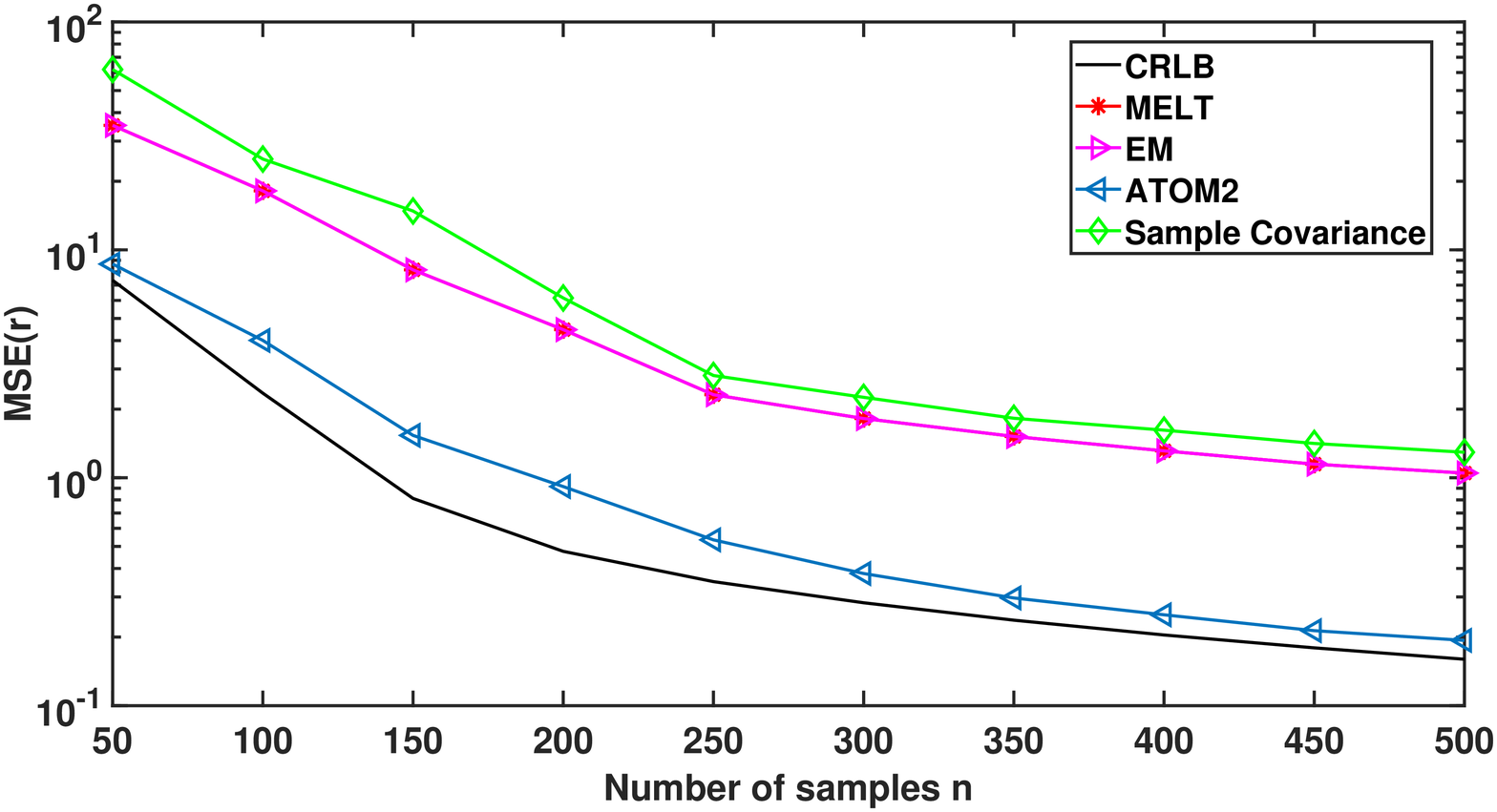}
\caption{MSE($\br$) vs. number of samples $n$ for TBT covariance matrix}
\label{tbt}
\end{figure}

ATOM2 is compared with the counterparts when the covariance matrix is constrained to be a TBT matrix. The ground truth covariance matrix of dimension $m=16$ with block size equal to $4$ is constructed by alternately projecting a  random Hermitian matrix onto the set of TBT matrices and the set of PSD matrices. Using (\ref{data}), the data $\by_{k}$ are generated using the ground truth TBT covariance matrix. The number of samples $n$ ranges between $50$ and $500$ in steps of $50$. The number of Monte-Carlo experiments is the same as in the previous simulation and the algorithms are compared in terms of the MSE metric and with CRLB as benchmark. Fig. \ref{tbt} shows that ATOM2 uniformly achieves the least MSE. As already highlighted the superior performance of the proposed method stems from the design criterion which does not require reparametrizing the covariance matrix using the CE. 

\subsection{MSE vs $n$ for Toeplitz covariance matrix with condition number constraint}\label{sec:5e}
\begin{figure}[h]
\centering
\begin{subfigure}[c]{0.45\textwidth}
\centering
\includegraphics[height=2.5in,width=3.0in]{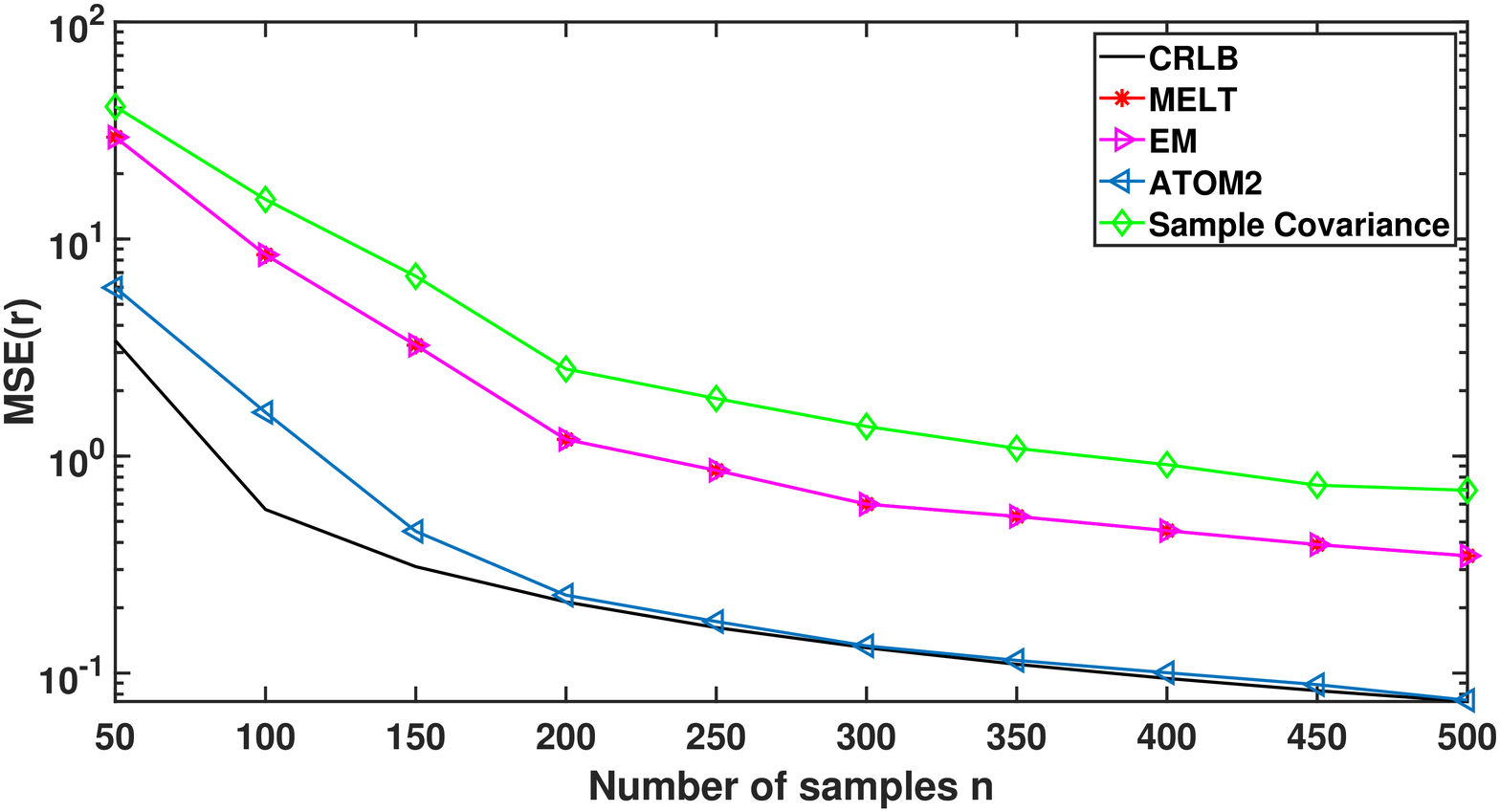}
\caption{}
\end{subfigure}
\hspace{5mm}
\begin{subfigure}[c]{0.45\textwidth}
\centering
\includegraphics[height=2.5in,width=3.0in]{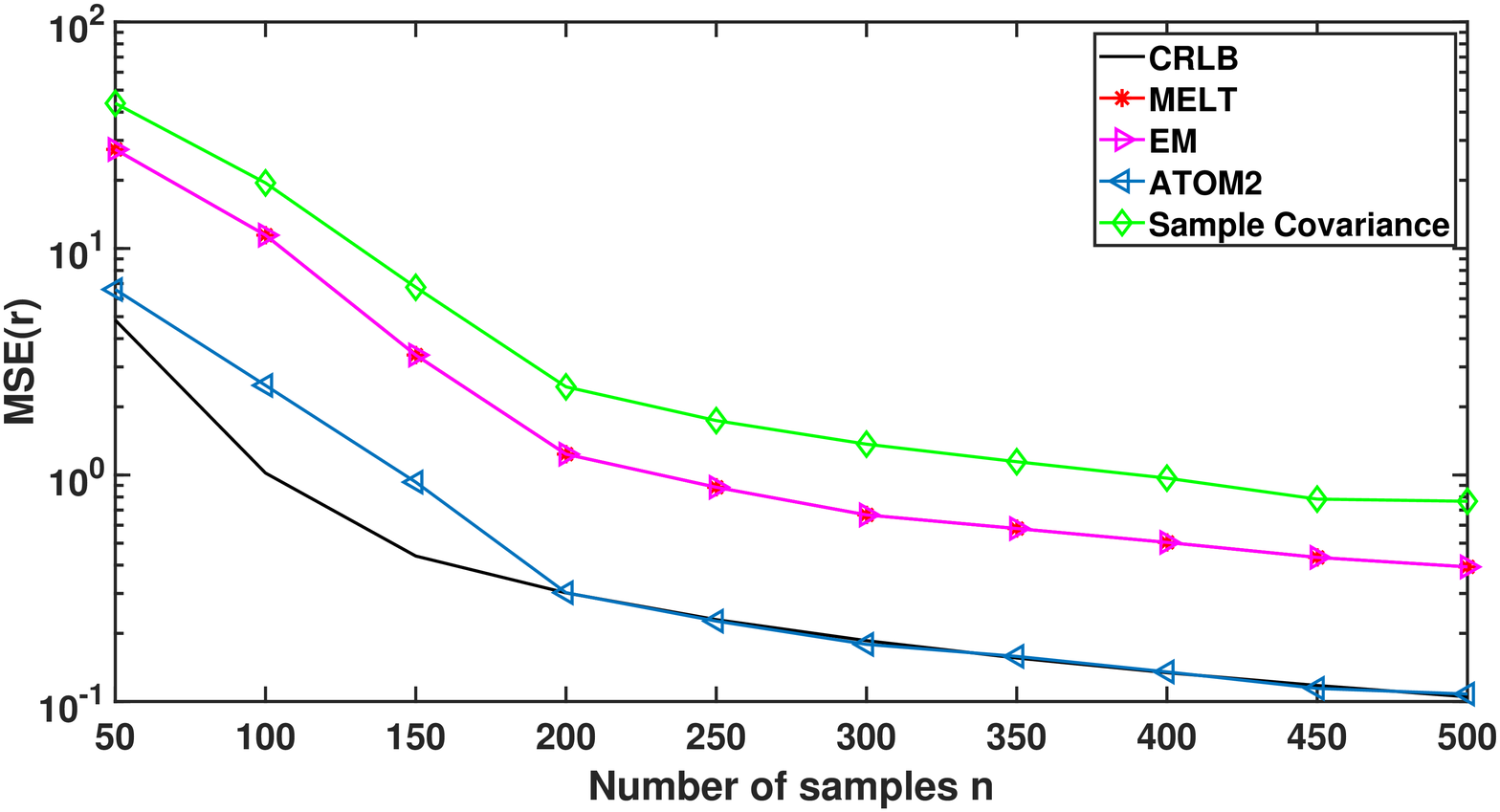}
\caption{}
\end{subfigure}
\caption{MSE($\br$) vs. number of samples $n$ for Toepltiz covariance matrix with condition number constraint. a) on-grid frequencies; b) off-grid frequencies}
\label{condiexp2}
\end{figure}

ATOM2 is compared with the state-of-the-art algorithms when the Toeplitz covariance matrix is forced to comply with a condition number constraint. The data $\by_{k}$ and ground truth Toeplitz covariance matrix of dimension $m=15$ are generated using the Fourier frequencies mentioned in Subsection \ref{sec:4b}. The condition number $\kappa$ is set equal to the condition number of the true underlying Toeplitz covariance matrix \cite{Augusto}. Fig. \ref{condiexp2}.a shows the MSE versus the number of samples $n$ for the on-grid frequency case study. Inspection of the figure highlights that the proposed algorithm, capitalizing on the a-priori knowldge on the covariance condition number, outperforms the state-of-the-art counterparts even when the frequencies lie on the Fourier grid. In Fig. \ref{condiexp2}.b one of the Fourier frequencies is replaced with an off-grid point $0.5$ rad/sec. Also in this last case ATOM2 yields the lowest MSE getting closer and closer to the CRLB.

\subsection{Radar Application}\label{sec:5f}
\begin{figure}[h]
\centering
\begin{subfigure}[c]{0.45\textwidth}
\centering
\includegraphics[height=2.5in,width=3.0in]{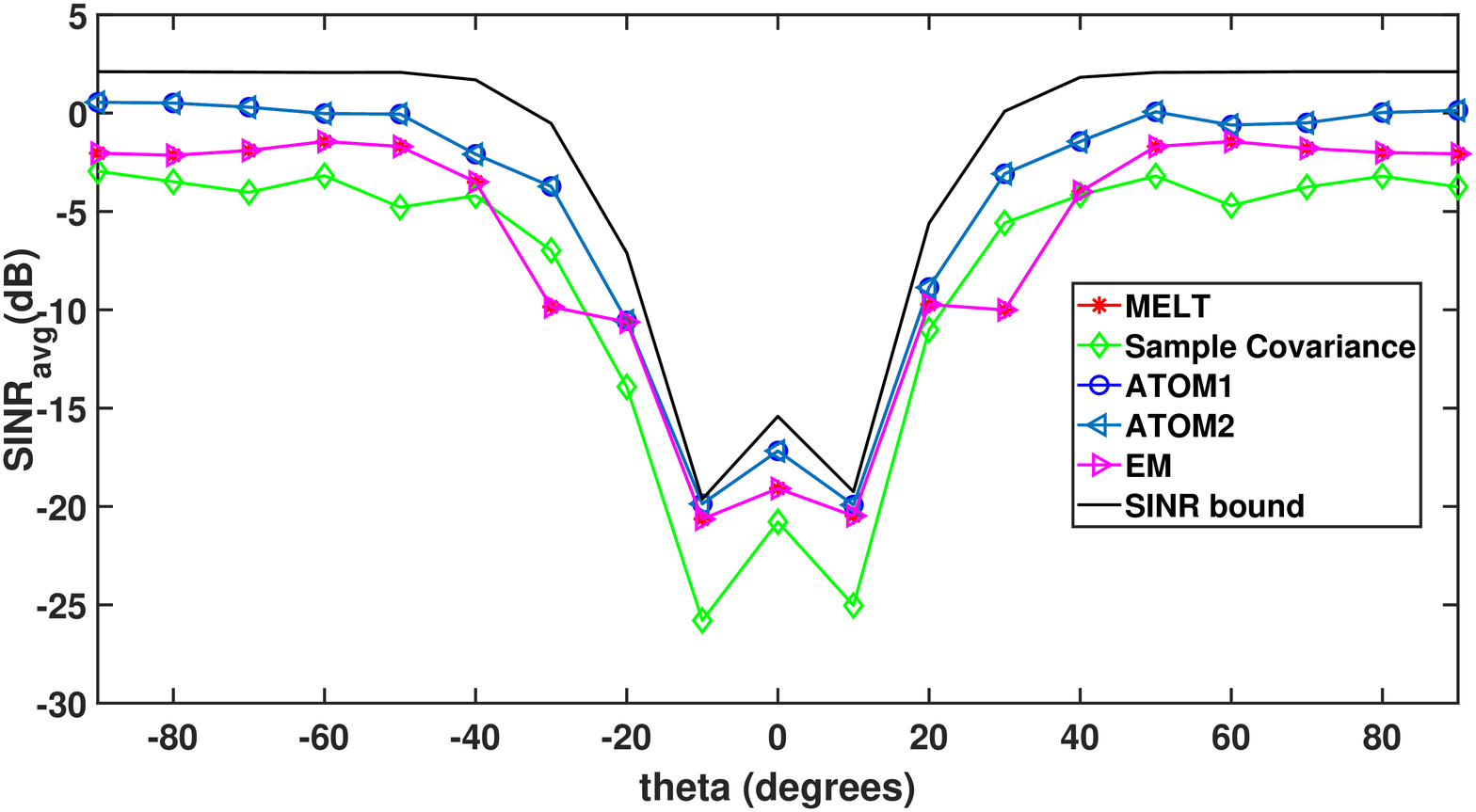}
\caption{}
\end{subfigure}
\hspace{5mm}
\begin{subfigure}[c]{0.45\textwidth}
\centering
\includegraphics[height=2.5in,width=3.0in]{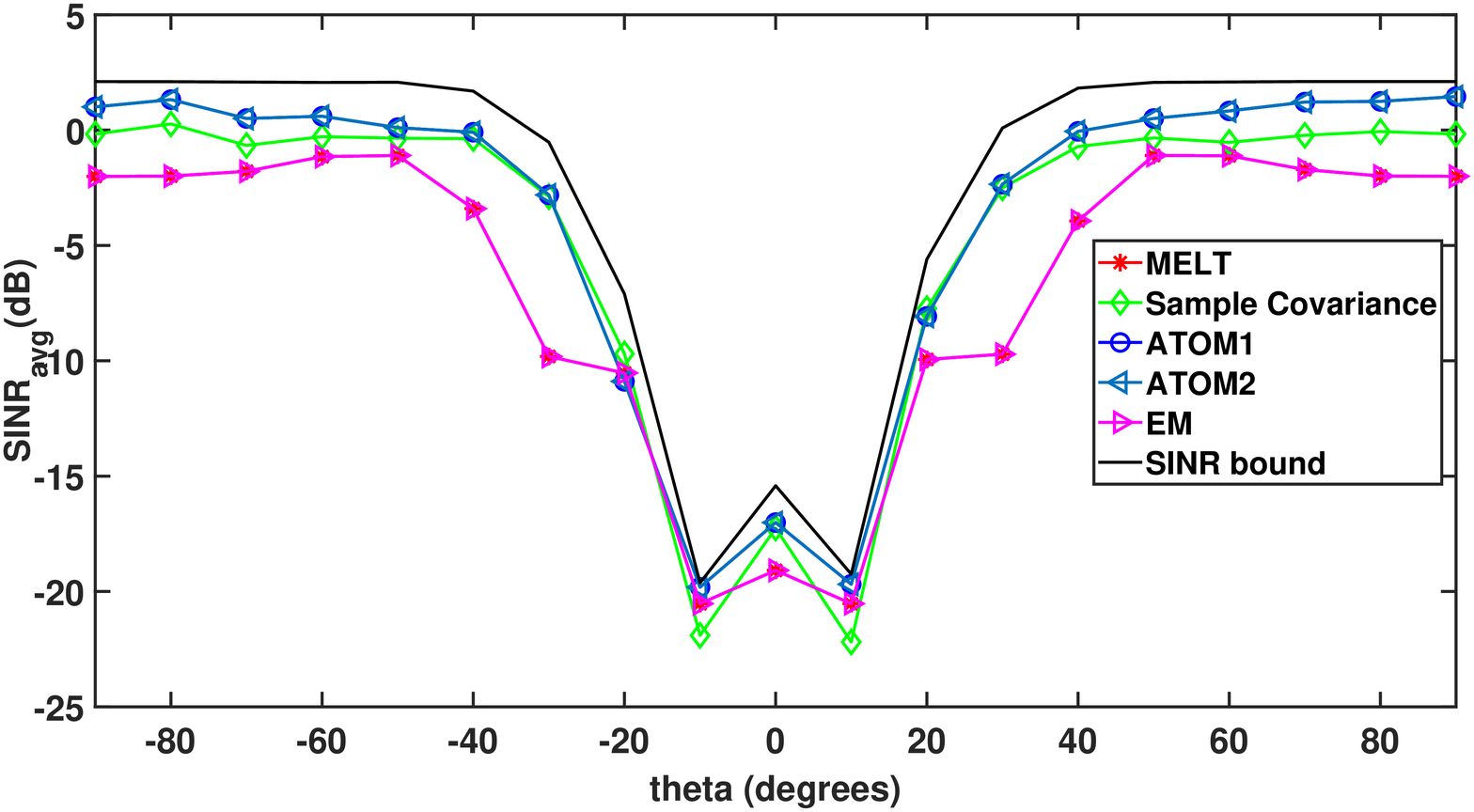}
\caption{}
\end{subfigure}
\begin{subfigure}[c]{0.45\textwidth}
\centering
\includegraphics[height=2.5in,width=3.0in]{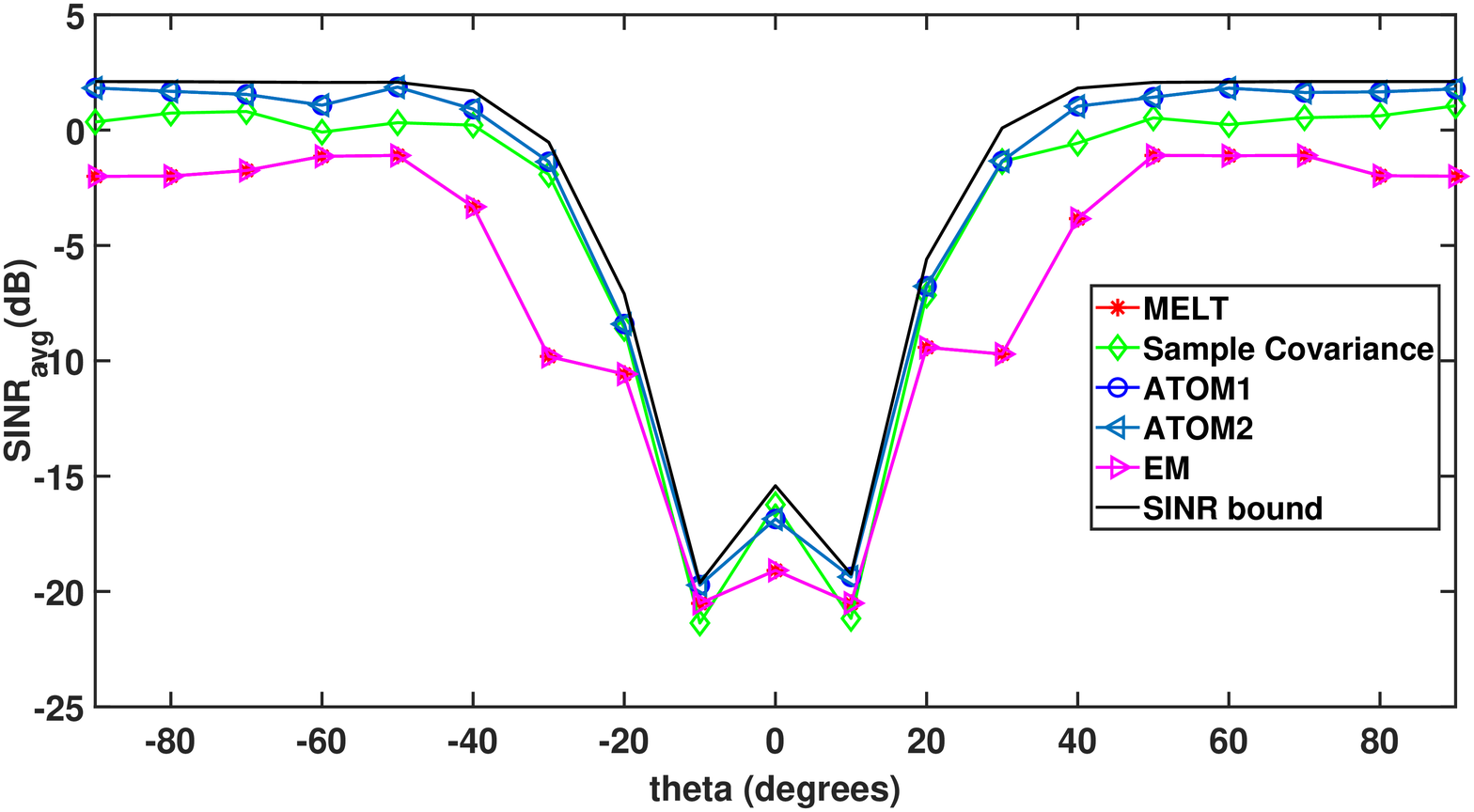}
\caption{}
\end{subfigure}
\caption{Average SINR vs $\theta$ in the presence of multiple jammers. a) $m=n $ b) $n=2m$, and c) $n=3m$}
\label{radar}
\end{figure}

In this subsection, the performance of the covariance estimation algorithms is evaluated with reference to the maximum achievable SINR in a radar spatial processing context. To this end, let us consider a radar system having uniform linear array with $m=6$ sensors, pointing toward the boresight direction. The distance between each sensors is equal to $d=\dfrac{\lambda}{2}$, where $\lambda$ is the radar operating wavelength. The interference covariance matrix is constructed as $\bR = \bR_{s} + \sigma_{a}^{2}\bI$ where $\sigma_{a}^{2}$ is the power level of the white disturbance noise and $\bR_{s}$ is obtained by summing the covariance matrices of the $J$ wide-band jammers i.e., 
\begin{equation}
    \begin{array}{ll}
    \bR_{s}(p,q)  = \displaystyle\sum_{i=1}^{J}\sigma_{i}^{2}\textrm{sinc}(0.5B_{f}(p-q)\phi_{i})e^{{j}(p-q)\phi_{i}} \hspace{1cm}  p,q = 1,2, \cdots m
    \end{array}
\end{equation}
where $\sigma_{i}^{2}$ denotes the power of the $i$-th interferer, $B_{f}=\dfrac{B}{f}$ is the fractional bandwidth, $B$ is the instantaneous Bandwidth, $f = \dfrac{c}{\lambda}$ and $c$ is the speed of light. The $i^{th}$ jammer phase angle is denoted by $\phi_{i}$ and is equal to $\phi_{i} = \dfrac{2\pi d \sin(\theta_{i})}{\lambda}$. The covariance estimation algorithms are compared using the average SINR metric
\begin{equation}
    {\rm{SINR}}_{\rm{avg}}= \dfrac{1}{K}\displaystyle\sum_{i=1}^{K}\dfrac{|\hat{\bw_{i}}^{H}\bs(\theta)|^{2}}{\hat{\bw}_{i}^{H}\bR\hat{\bw_{i}}}
\end{equation}
where $K$ is the number of Monte-Carlo trials, $s(\theta)$ is the $m$ dimensional steering vector and ${\hat{\bw}_{i}} = {\hat{\bR}_{i}}^{-1}\bs(\theta)$ is the estimate of the optimal weight vector for spatial processing with ${\hat{\bR}_{i}}$ being the estimate of the interference covariance matrix for the $i$-th trial. Previous assumptions imply that for the above considered linear array, the steering vector is given by $s(\theta)= [1, e^{j\pi\sin(\theta)},\cdots, e^{j\pi\sin(\theta)(m-1)}]^{T}$. The simulation setup assumes $J=2$ jammers having the same power $\sigma_{i}^{2}= 20$ dB with $\theta_{1}=9.8^{\circ}$ and $\theta_{2}=-8.8^{\circ}$. The fractional bandwidth of the two jammers and white noise power level is equal to 0.3 and $10$ dB, respectively. The average SINR as a function of theta is shown in Fig. \ref{radar}.a,  Fig. \ref{radar}.b, and  Fig. \ref{radar}.c  for $n=m$, $n=2m$, and $n= 3m$, respectively. These figures also indicate the SINR bound calculated as $\bs(\theta)^{H}\bR^{-1}\bs(\theta)$. Inspection of the plots highlights that as the number of samples $n$ increases ATOM1 and ATOM2 gets nearer and nearer to the SINR bound. 
\section{Conclusion}\label{sec:5}
In this paper, MLE of TSC matrices is considered. Precisely leveraging the MM framework, two iterative algorithms ATOM1 and ATOM2 are developed. Both inherit the key properties of MM i.e., they monotonically decrease the underlying cost function with guaranteed convergence to a stationary point of the equivalent MLE problem. Subsequently, ATOM2 is extended to handle covariance matrix MLE forcing other Toeplitz-related structures, such as: banded Toeplitz, TBT, low rank Toeplitz plus a scalar matrix, and Toeplitz structure satisfying a condition number constraint. Simulation results indicate that the proposed algorithms can perform better than some state-of-the-art techniques in terms of MSE and the SINR metrics. \\
Some of the possible future research directions are now outlined. In the present work, for low rank TSC matrix estimation, the rank ($r$) of the covariance matrix is assumed to be known. However, in many scenarios a perfect prior knowledge of $r$ could not be available. Therefore, the proposed algorithm could be extended and analyzed to the case of unknown $r$. Another possible extension of the proposed technique could be MLE of a Toeplitz covariance matrix assuming a compound Gaussian distribution for the underlining data which has a significant application in low-grazing angle target detection \cite{appp,gini}.  
\section*{Appendix A}\label{appendixA}
\vspace{-2mm}
\section*{Equivalence of (\ref{n0}) and (\ref{n1})}
Let us cast Problem (\ref{n0}) as
\begin{equation}\label{newo}
\begin{array}{ll}
\underset{{\bR}  \in Toep, \bR \succ 0, q>0}{\rm minimize} \: q+ \log|\bR| \:\:
{\rm s.t.}\quad \dfrac{1}{n}\displaystyle\sum_{i=1}^{n}\by_{i}^{H}{\bR}^{-1}\by_{i} \leq q
\end{array}
\end{equation}
Now, with variable substitution $\bX= \bR q$, Problem (\ref{newo}) is equivalent to
\begin{equation}\label{neweq}
\begin{array}{ll}
\underset{{\bX}  \in Toep,\bX \succ 0, q>0}{\rm min} \: q+ \log|\dfrac{\bI}{q}| +\log|\bX| \:\:
{\rm s.t.}\quad \dfrac{1}{n}\displaystyle\sum_{i=1}^{n}\by_{i}^{H}{\bX}^{-1}\by_{i} \leq 1
\end{array}
\end{equation}
The minimizer w.r.t. $q$ is constant and is given as $q^{*}= m$. Next, an optimal solution w.r.t. $\bX$ can be obtained by solving the following optimization problem
\begin{equation}
\begin{array}{ll}\tag{$P_{2}$}
\bX^{*}=\underset{{\bX}  \in Toep, \bX \succ 0}{\rm arg\:min} \: \log|\bX| \:\:
{\rm s.t.}\quad \dfrac{1}{n}\displaystyle\sum_{i=1}^{n}\by_{i}^{H}{\bX}^{-1}\by_{i} \leq 1
\end{array}
\end{equation}
As a result, a minimizer to Problem (\ref{n0}) is $\bR^{*}=\dfrac{\bX^{*}}{q^{*}} = \dfrac{\bX^{*}}{m} $. 
\section*{Appendix B}\label{appendixB}
\vspace{-2mm}
\section*{SDP Formulation of Problem (\ref{s1})}
Problem (\ref{s1}) can be written as 
\begin{equation}\label{s2}
\begin{array}{ll}
 \underset{{\bX\in Toep,\bX \succ 0}}{\rm minimize} \: \textrm{Tr}\left(({\bX}_{t})^{-1}{\bX}\right)\\
{\rm subject\ to}\quad \bar{\by}^{H}\boldsymbol{\hat{X}}^{-1}\bar{\by}\leq {n} \\
\quad\quad \quad \quad \quad \boldsymbol{\hat{X}}=\bI \otimes \bX
\end{array}
\end{equation}
where $\boldsymbol{\hat{X}}$ is a block diagonal matrix with the diagonal blocks equal to $\bX$ and $\bar{\by}$ is a vector obtained by stacking the vectors $\by_{1},\by_{2},\cdots,\by_{n}$. Being $\bX \succ 0$, using the Schur complement theorem \cite{boyd}, Problem (\ref{s2}) can be formulated as an SDP
\begin{equation}
    \begin{array}{ll}
  \underset{{\bX\in Toep,\hatbX}}{\rm minimize} \: \textrm{Tr}\left(({\bX}_{t})^{-1}{\bX}\right)\\
{\rm subject\ to} \:  \begin{bmatrix}    
   n &  \bar{\by}^{H}\\
   \bar{\by} & \boldsymbol{\hat{X}}
   \end{bmatrix} \succeq 0
    \end{array}
\end{equation}
Once again leveraging the Schur complement theorem, the following optimization problem is obtained
\begin{equation} 
\begin{array}{ll}
 \underset{{\bX\in Toep,\hatbX}}{\rm minimize} \: \textrm{Tr}\left(({\bX}_{t})^{-1}{\bX}\right)\\
{\rm subject\ to}\quad \boldsymbol{\hat{X}} \succeq \dfrac{1}{n} \bar{\by}\bar{\by}^{H}
\end{array}
\end{equation}
which is also an SDP.
\section*{Appendix C}\label{appendixC}
\section*{Proof of Lemma \ref{lemma 3}}
To begin with, let us denote by $g(\bX|\bX_{t})$ the objective function involved in the surrogate optimization problem exploited by either ATOM1 or ATOM2. This function, regardless of the method, satisfies the following two inequalities
\begin{equation}\label{d1}
g(\bX_{t}|\bX_{t}) = \tilde{f}(\bX_{t})
\end{equation}
\begin{equation}\label{d2}
g(\bX_{t+1}|\bX_{t}) \geq g\tilde{f}(\bX_{t+1})
\end{equation}
where $\tilde{f}(\bX)=\log|\bX|$. Leveraging the above inequalities, it follows that
\begin{equation}\label{d3}
\tilde{f}(\bX_{t+1})  \overset{(a)} \leq g(\bX_{t+1}|\bX_{t})  \overset{(b)} \leq g(\bX_{t}|\bX_{t})  \overset{(c)}= \tilde{f}(\bX_{t})
\end{equation}
In (\ref{d3}), the inequality (a) and equality (c) stem from (\ref{d2}) and (\ref{d1}), respectively; besides, the inequality (b) is obtained by exploiting the fact that ATOM1 and ATOM2 globally solve the corresponding convex surrogate optimization problem. Therefore, (\ref{d3}) implies that the sequence of objective value of Problem $(P_{2})$ generated by the proposed algorithms is monotonically decreasing , i.e.,
\begin{equation}\label{d4}
\tilde{f}(\bX_{0}) \geq \tilde{f}(\bX_{1}) \geq \tilde{f}(\bX_{2}) \geq \cdots
\end{equation} 
Next, let us denote by $\bZ$ a cluster point to $\{\bX_{t}\}$ and let $\{\bX_{r_{t}}\}$ be a subsequence  of $\{\bX_{t}\}$ converging to $\bZ$. Then, from (\ref{d1}), (\ref{d2}), and (\ref{d4}) 
\begin{equation}
\begin{array}{ll}
g\left(\bX_{r_{t+1}}|\bX_{r_{t+1}}\right)= \tilde{f}\left(\bX_{t_{j+1}}\right) \leq \tilde{f}\left(\bX_{r_{t}+1}\right) \leq 
g\left(\bX_{r_{t}+1}|\bX_{r_{t}}\right)\leq g\left(\bX|\bX_{r_{t}}\right), \forall\,\, {\mbox{feasible}}\,\,\bX.
\end{array}
\end{equation}
Thus, letting $t \rightarrow \infty$
\begin{equation}
g(\bZ|\bZ) \leq g(\bX|\bZ),
\end{equation}
which implies that  $g'(\bZ|\bZ;\bD) \geq 0$ where $g'(\cdot|\bZ;\bD)$  is the directional derivative of the surrogate function at point $\bZ$ in a feasible direction $\bD$. Finally, by Proposition 1 in \cite{conv}, the surrogate function $g(\cdot|\bZ)$ and  the objective function $\tilde{f}(\cdot)$ have the same first order behavior at $\bZ$. Therefore, $g'(\bZ|\bZ;\bD) \geq 0$ implies that  $\tilde{f}'(\bZ; \bD) \geq 0$. Hence, $\bZ$ is a stationary
point of the objective function $\tilde{f}(\cdot)$ of $(P_{2})$.
\bibliographystyle{IEEEtran} 
\bibliography{refs}
\end{document}